\documentclass[a4paper,12pt]{article}
\usepackage[dvips]{graphics}
\usepackage{epsfig}
\usepackage{psfrag}
\usepackage[psamsfonts]{amssymb}
\usepackage{amsmath}
\usepackage{indentfirst}
\usepackage{amssymb}
\usepackage{cite}
\newtheorem{proposition}{Proposition}
\newtheorem{remark}{Remark}
\newtheorem{definition}{Definition}
\newtheorem{conjecture}{Conjecture}
\newtheorem{property}{Property}

\begin{document}

\title{Statistics of seismic cluster durations}

\author{A. Saichev$^*$ and D. Sornette\\
ETH Zurich\\
Department of Management, Technology and Economics\\
Scheuchzerstrasse 7, 8092, Zurich, Switzerland\\
dsornette@ethz.ch\\
$^*$deceased 8 June 2013}

\date{\today}

\maketitle

\begin{abstract}
Using the standard ETAS model of triggered seismicity, we present
a rigorous theoretical analysis of the main statistical properties of
temporal clusters, defined as the group of events triggered by a given main shock of fixed
magnitude $m$ that occurred at the origin of time, at times larger than some present time $t$.
Using the technology of generating probability function (GPF), 
we derive the explicit expressions for the GPF of the number of future offsprings 
in a given temporal seismic cluster,  defining, in particular, the statistics of the cluster's duration 
and the cluster's offsprings maximal magnitudes.  
We find the remarkable result that the
magnitude difference between the largest and second largest event in the
future temporal cluster is distributed according to the regular Gutenberg-Richer law
that controls the unconditional distribution of earthquake magnitudes.
For earthquakes obeying the Omori-Utsu law
for the distribution of waiting times between triggering and triggered events, 
we show that the distribution of the durations of temporal clusters of events of 
magnitudes above some detection threshold $\nu$ has a power law tail that is fatter
in the non-critical regime $n<1$ than in the critical case $n=1$. This paradoxical
behavior can be rationalised from the fact that generations of all orders cascade very fast
in the critical regime and accelerate the temporal decay of the cluster dynamics.
\end{abstract}

\section{Introduction}
 
The present article is the last in a series  \cite{SHJGR2002, HSorJGR2003, SaiSor2004, SaiSorBath2005, SaiHSor2005, SorWern2005, SaiSorVJ2005, SaiSor2006, SaiSorRenorm2006, SaiSorPRL2006, SaiSor2007, SaiSorTimes2007, SorUtSai2008, WerSor2008, SaiSor2011} devoted to the analysis of various statistical properties of the aftershocks triggering process
of the ETAS model \cite{SH,KK,KK2,Ogata,Ogata2}), which serves as a standard benchmark in statistical seismology.
Here, we present derivations and results concerning statistical properties of the temporal seismic clusters, 
made of future events triggered after the current time $t$ by some main shock of magnitude $m$ that 
occurred as some previous instant taken as the origin of time.

We mostly investigate the statistical properties of the durations of temporal seismic clusters and of the maximal magnitude 
over all events in the clusters. Specifically, we derive the dependence of the probability density function (pdf) of the 
duration of seismic clusters as a function of the main shock magnitude $m$, of a magnitude threshold $\nu$ 
of earthquake detection and of the branching ratio $n$. In addition, we study the characteristic properties of the maximal magnitudes of future offsprings triggered after the current time $t$.

The results obtained here have a broader domain of application than statistical seismology,
and can be used for any system that can be described by the class of self-excited conditional 
Poisson process of the Hawkes family \cite{Hawkes1,Hawkes2,Hawkes3} 
and its extension and more generally to branching processes.
For instance, some of the stochastic processes in financial markets can be well represented by
this class of models, for which the triggering and branching processes capture the herding nature
of market participants. The Hawkes process has been successfully involved in issues as diverse as estimating
the volatility at the level of transaction data, estimating the market stability \cite{FiliSor12,FiliSor15},
accounting for systemic risk contagion, devising optimal execution strategies or
capturing the dynamics of the full order book \cite{Bacry1}.

The article is organised as follows. Section~2 presents the formulation of the version of the
Hawkes model, known at the ETAS model of triggered seismicity, with particular emphasis
on statistics of temporal seismic clusters statistics. Section 2 also introduces the fractional exponential model 
as a convenient parameterisation of the distribution of waiting times between triggering and triggered events
(also known as the bare Omori law). The fractional exponential model provides a rather accurate
approximation of the well-known modified Omori-Utsu law.
Section~3 derives the explicit and approximate expressions for the generating probability function  
(GPF) of the number of future offsprings in a given temporal seismic cluster, 
defining, in particular, the statistics of the cluster's duration and the cluster's offsprings maximal magnitudes.
Section~4 presents a detailed statistical analysis of the seismic cluster's duration statistics and the statistics of the maximal offsprings magnitude. It is shown in particular that, in the subcritical case, one may use, without significant error, the
so-called one-daughter approximation in which each event can trigger not more than one first-generation aftershock.
All proofs of our four main results, presented in the form of four propositions, are given in appendices.
Section 5 conclude.

\section{Statistical description of the future offsprings in the framework of the ETAS model}

Let us consider a main shock occurring at time $t=0$ with magnitude $m$. 
To make precise our investigation of the statistical properties of the aftershocks triggered by
the main shock (consisting of the main shock's direct aftershocks, the direct aftershocks 
of the first generation aftershocks and so on), we use the ETAS model \cite{Ogata,KK,KK2}, whose main assumption
is that all earthquakes obey the same laws governing the generation of triggered earthquakes.
Each earthquake is thus potentially the ``mother'' of triggered events, which themselves can trigger their
own ``daughters'' and so on. In the ETAS model, there are two categories of earthquakes:
(i)  the main shocks that are supposed to be ``immigrants'', i.e. they are not triggered by previous earthquakes,
and (ii) all the other earthquakes that are triggered by some previous event, be it a main shock or one of the
event it has triggered either directly or indirectly through a cascade.



\subsection{The ETAS model and its laws}

The cornerstones of the ETAS model are based on three well-known statistical laws that govern
the process of earthquake triggering.
The following subsections \ref{yhyg}-\ref{wrnb3hg} enunciate the four fundamental definitions of the
ETAS model, which describe the properties shared by all earthquakes.

\subsubsection{The Gutenberg-Richter law in the ETAS model \label{yhyg}}

The well-known \emph{Gutenberg-Richter law} (GR) states that earthquakes occur with magnitudes
distributed according to the complementary cumulative distribution function 
\begin{equation}\label{grldef}
p(m) := \text{Pr}\left\{m' \geqslant m\right\} = 10^{-b m}~,
\end{equation}
where the $b$-value is often found empirically close to $1$ \cite{YavorHiemer2015}.

In the ETAS model, the GR law is assumed to apply both for the main shocks and for their aftershocks,
as well as for subsequent aftershocks of aftershocks over all generations. Moreover, the ETAS model
posits that the magnitudes are independent random variables, i.e. there is no (unconditional) dependence between
the magnitudes of any earthquake in a given seismic catalog. The magnitudes are thus i.i.d. random variables
distributed according to (\ref{grldef}).

\subsubsection{The fertility law in the ETAS model \label{ethn3hg}}

The \emph{fertility law} states that the mean number $\overline{R}_d(m)$ of direct (first generation) aftershocks triggered by 
a given earthquake of magnitude $m$ is exponentially large in the magnitude of the mother earthquake \cite{Helmalpha03}:
\begin{equation}\label{fertlaw}
\overline{R}_d(m) = \kappa \mu , \qquad \mu := 10^{\alpha m} ~,
\end{equation}
where $\alpha$ is in general found to be smaller than $b$, with typical values close to $0.8$.
The parameters $\kappa$ as well as $\alpha$ may depend on regional properties of seismicity.

For simplicity of notations, we assume that all magnitudes are positive and all events with
a positive magnitude has the ability to trigger future events. In the standard ETAS model, 
one introduces a characteristic cut-off magnitude $m_0$, below which events are sterile, i.e.
do not trigger other events. This cut-off magnitude is needed to ensure that the ETAS model
is well-defined, otherwise, the swarms of arbitrary small earthquakes make the seismic
activity divergent and ill-defined \cite{SorWern2005}. Our parameterisation thus
amounts to take $m_0=0$, which is nothing but a translation in the magnitude scale
that has no impact on the calculations and results.

\subsubsection{The modified Omori-Utsu law in the ETAS model \label{rjymjh}}

The \emph{modified Omori-Utsu law} specifies the distribution $f(t)$ of waiting times $\{T_k\}$ between 
a mother earthquake and its direct (first-generation) offsprings \cite{{Utsuomori95}}:
\begin{equation}\label{domoridist}
f(t) = \frac{\theta c^\theta}{(c+t)^{\theta+1}} , \qquad \theta\in (0,1) , \qquad c>0 .
\end{equation}
By construction, it gives the dependence of the
rate of first-generation aftershocks as a function of time $t$ counted since the mother earthquake.

The ETAS model assumes further that the modified Omori-Utsu law applies for all earthquakes, whatever
their rank in the generation ordering. Thus, all earthquakes have the potential to trigger their aftershocks
with delays given by expression (\ref{domoridist}).

\subsubsection{Poisson statistics in the ETAS model \label{wrnb3hg}}

Combining the assumptions stated in subsections \ref{yhyg}-\ref{rjymjh} that all earthquakes
are treated equally in the sense that they all possess the same propensity for triggering earthquakes
with the same time dependence given by the Omori-Utsu law and with i.i.d. magnitudes, 
it derives that the total number $R_d(m)$ of first-generation daughters triggered by the mainshock of magnitude $m$ obeys 
the Poisson statistics. This means that the probability $q_d(r|m)$ that the number $R_d(m)$ of 
first-generation daughters takes the value $r$ is given by
\begin{equation}\label{rdempois}
q_d(r|m) := \text{Pr}\left\{R_d(m) = r\right\} = \frac{(\kappa \mu)^r}{r!} ~ e^{-\kappa \mu} .
\end{equation}
Accordingly, the Generating Probability Function (GPF) of the total number of first-generation daughters is given by 
\begin{equation}\label{Poisson}
G_d(z|m) := \mathbb{E}\left[z^{R_d(m)} \right] = \sum_{r=0}^\infty q_d(r|m) \, z^r = e^{\kappa \mu (z-1)} .
\end{equation}

As a consequence of the number $R_d(m)$ of first-generation daughters 
obeying Poissonian statistics \eqref{Poisson}, and given that their occurrence times $\{T_k\}$
are statistical independent, we can state that
\begin{proposition}\label{propositionone}
\textnormal{
Given a fixed observation time $t$, the random numbers of first-generation daughters
of the main shock that occurred in the past (before $t$) and that will come in the future (after $t$)
are statistically independent and obey to the Poissonian statistics.
}
\end{proposition}
The proof of Proposition 1 is given in the Appendix.

\subsection{Statistics of the offsprings magnitudes in the framework of the algebraic GPF approximation}

For the known and fixed main shock magnitude $m$, the GPF of the number of its first-generation daughters
is given by expression \eqref{Poisson}. In contrast, the GPF of the number of first-generation events
of an arbitrary first-generation daughter (i.e. the number of grand-daughters of the main shock via the filiation of one
of its daughters) is given by the average of \eqref{Poisson} over the GR distribution of magnitudes  \eqref{grldef},
since the first-generation daughters have random iid magnitudes:
\begin{equation}
G_d(z) = - \int_0^\infty G_d(z|m) d p(m)~ .
\end{equation}
This yields
~\begin{equation}\label{thetoneexpr}
G_d(z) = \gamma \kappa^\gamma (1-z)^\gamma \Gamma(-\gamma, \kappa (1-z)) , \qquad \gamma = b\big/ \alpha ~,
\end{equation}
where $\Gamma(a,z)$ is the incomplete gamma function. 

For real aftershock sequences, the parameter $\gamma$ belongs to the interval: $\gamma \in (1,2)$
(see more detailed discussion in \cite{SaiHSor2005}). In particular, for $(b = 1, \alpha = 0.8)$, we have $\gamma = 1.25$.
For such a value, for the convenience of future analytical derivations,
we may replace the exact expression \eqref{thetoneexpr} by the first three power terms of its Taylor expansion 
in the variable $(1-z)$:
\begin{equation}\label{gonetwo}
\mathcal{G}_d(z) : = 1- n (1-z) + \rho (1-z)^\gamma , \qquad 0<\gamma<2 ~,
\end{equation}
where
\begin{equation}\label{enrhodef}
n := \frac{\kappa \gamma}{\gamma-1} , \qquad \rho = \kappa^\gamma \gamma \Gamma(-\gamma) .
\end{equation}
Parameter $n$ is the so-called ``branching ratio'', which quantifies how many first-generation daughters
are triggered per mother. A value close to $1$ corresponds to the approach to the critical regime \cite{SHJGR2002}.
$n$ has also the meaning of being the average fraction of triggered events in the whole population \cite{HSorGRL2003}.
Parameter $n$ will play a crucial role in the following analysis of the statistical properties of seismic clusters, in particular
in the two fundamental regimes, the subcritical ($n<1$) and critical ($n=1$) regimes, which are relevant
to real seismic activity.

\begin{figure}
\begin{center}
\includegraphics[width=0.75\linewidth]{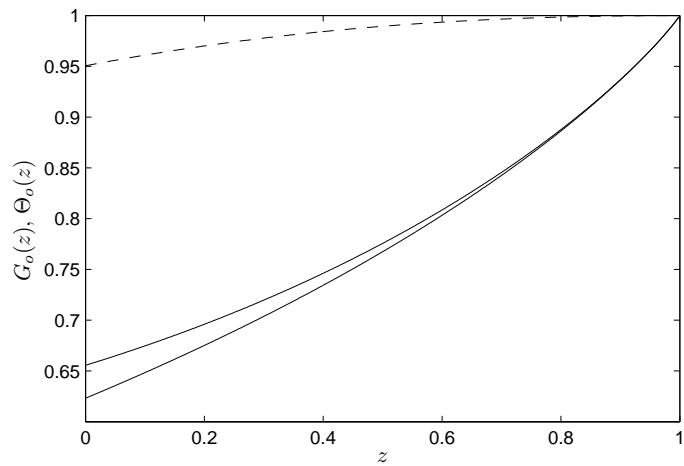}\\
\end{center}
\caption{The lower solid line shows the exact GPF $G_d(z)$ \eqref{thetoneexpr} as a function of the variable $z$. The upper solid line 
is its algebraic approximation $\mathcal{G}_d(z)$ \eqref{gonetwo}. The dashed line is the ratio of the former to the later,
which quantifies the accuracy of the algebraic approximation. The parameters are $b=1$, $\alpha=0.8$ and $\kappa=0.2$, 
corresponding to $\gamma=1.25$ and $n=1$. One can notice that the algebraic approximation becomes
excellent as $1-z \to 0$, and the two expressions are undistinguishable for $1-z < 0.2$.}\label{gpfalg}
\end{figure}

Plots of the GPF $G_d(z)$ \eqref{thetoneexpr} and its algebraic approximation $\mathcal{G}_d(z)$ \eqref{gonetwo} for the typical values $(b = 1, \alpha = 0.8)$ and for $\kappa=0.2$ (that is, for $\gamma=1.25$ and $n=1$) are depicted in figure~\ref{gpfalg}.

Parameter $\gamma$ controls the asymptotic decay for $r\to\infty$ of the probability
\begin{equation}
q_d(r) = \frac{1}{r!} \frac{d^r G_d(z)}{dz^r}\bigg|_{z=0}
\end{equation}
that an arbitrary daughter has $r$ first-generation offsprings. One can show (see, for instance, \cite{SaiHSor2005}),
using either the exact $G_d(z)$ \eqref{thetoneexpr} and its algebraic 
approximation $\mathcal{G}_d(z)$ \eqref{gonetwo}, that $q_d(r)$ has the following asymptotic tail
\begin{equation}\label{qudelar}
q_d(r) \simeq  \frac{\rho}{\Gamma(-\gamma)}\cdot {1 \over r^{1+\gamma}} , \qquad r\gg 1 ~.
\end{equation}
Since $1 \leq \gamma <2$, the mean number of descendants of first-generation exists,  but not its variance.

\subsection{The modified Omori-Utsu law and its fractional exponential approximation}

The modified Omori-Utsu law \eqref{domoridist} used in the ETAS specification follows Ogata's 
formulation of the ETAS model \cite{Ogata}. Earlier, Kagan and Knopoff \cite{KK,KK2} introduced a
version of ETAS (under a different name), which had essentially all its ingredients, except 
for the expression of the function $f(t)$ controlling the distribution of waiting times between
a mother earthquake and its first-generation offsprings. Kagan and Knopoff \cite{KK,KK2} 
used a pure power law $f(t) \sim 1/t^{1+\theta}$ truncated to zero for $t<c$ for some positive 
characteristic time $c$. The difference between Ogata's and 
Kagan and Knopoff's specifications of the memory kernel $f(t)$ amounts to a change 
in ``ultraviolet'' cut-off, which was shown in Ref.\cite{SS99} to have no
significant impact on the dynamics and overall generating process.

Here, we propose to use another ultra-violet cut-off, which has significant advantages 
for the analytical computations that we develop below, without significant 
impact on the main characteristic of the model, namely 
its heavy tail power law asymptotics
\begin{equation}\label{powou}
f(t) \sim t^{-\theta-1} , \qquad t \to \infty ~.
\end{equation}
We thus propose to use the so-called \emph{fractional exponential distribution}, which has the same power 
law asymptotics \eqref{powou}, as the modified Omori-Utsu law. 

To motivate the fractional exponential distribution, let us consider Laplace transform $\hat{f}(u)$ of
the probability density function (pdf) $f(t)$:
\begin{equation}
\hat{f}(u) := \int_0^\infty f(t) e^{-u t} dt~ .
\end{equation}
It is well-known that the two following power law asymptotics are equivalent:
\begin{equation}\label{efhatefas}
\begin{array}{c}
f(t) \simeq \frac{\epsilon}{\Gamma(-\theta)} \cdot t^{-\theta-1} , ~ t\to\infty \quad \Leftrightarrow \quad 1-\hat{f}(u) \simeq \epsilon \cdot u^\theta , ~ u\to 0 ~, ~{\rm for}~ \theta \in (0,1) .
\end{array}
\end{equation}

\begin{definition}
\textnormal{
The fractional exponential distribution, denoted $f_\theta(t)$, is defined by its Laplace transform
\begin{equation}\label{feqlim}
\hat{f}_\theta(u) := \frac{1}{1+ u^\theta} ~.
\end{equation}
}
\end{definition}

\begin{remark}
\textnormal{
The fractional exponential distribution, with Laplace image \eqref{feqlim}, corresponds to the canonical case where the factor $\epsilon$ in the above asymptotics \eqref{efhatefas} is equal to $1$. 
The modified Omori-Utsu law \eqref{domoridist} gives the same asymptotic power law 
with the correspondence 
\begin{equation}\label{ctgam}
c^\theta = \Gamma^{-1}(1-\theta) \qquad \Rightarrow \qquad c = \Gamma^{-1/\theta}(1-\theta) ~.
\end{equation}
}
\end{remark}

One can show that the fractional exponential distribution, defined by the Laplace image \eqref{feqlim}, is given by
\begin{equation}\label{hateftheta}
f_\theta(t) = t^{\theta-1} \cdot E_{\theta,\theta}(-t^\theta) , \qquad
\theta \in (0,1) ~,
\end{equation}
where $E_{\alpha,\beta}(z)$ is the generalized Mittag-Leffler function
\begin{equation}
E_{\alpha,\beta}(z) := \sum_{j=0}^\infty \frac{z^j}{\Gamma(\alpha j + \beta)}~ .
\label{etn4hw}
\end{equation}

The fractional exponential distribution is characterised by two power law asymptotics,
one for short times and the other for long times:
\begin{equation}\label{efthas}
f_\theta(t) \simeq
\begin{cases}\displaystyle
- \frac{t^{-\theta-1}}{\Gamma(-\theta)} , & t\gg 1 ,
\\[4mm] \displaystyle
\quad \frac{t^{\theta-1}}{\Gamma(\theta)} , & t\ll 1 ,
\end{cases}
\qquad \theta \in (0,1) .
\end{equation}
By construction, it has the same power law asymptotics as the modified Omori-Utsu law \eqref{domoridist} 
at long times, with the matching (\ref{ctgam}) of the prefactors as already mentioned.

\begin{figure}
\begin{center}
  \includegraphics[width=0.99\linewidth]{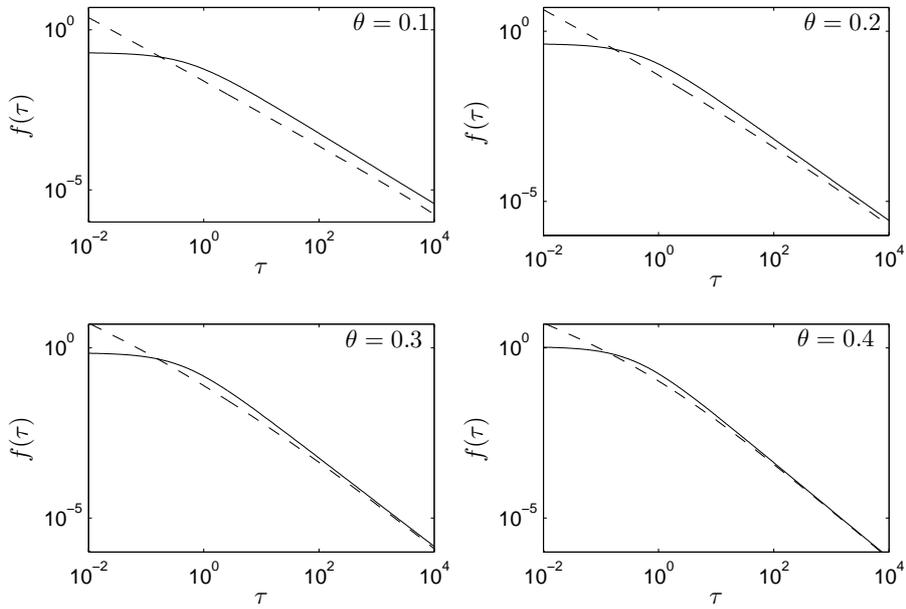}\\
\end{center}
  \caption{Plots in double logarithm scales, for four different $\theta$ values, of the modified Omori-Utsu law pdf $f(t)$ \eqref{domoridist} (solid lines) and the corresponding fractional exponential distributions $f_\theta(t)$ \eqref{hateftheta} (dashed lines). 
 To make the two pdf's comparable, we use the scale parameter $c$ given by \eqref{ctgam}.
 For $\theta=0.4$, one can observe best the transition from the slope $-1-\theta$ at large times to
 the slope $-1+\theta$ at small times.}\label{omfrac}
\end{figure}

Figure~\ref{omfrac} compares in log-log scales the modified Omori-Utsu law \eqref{domoridist} and the corresponding fractional exponential distribution $f_\theta(t)$ \eqref{hateftheta}, for four different values of the exponent $\theta$.
These plots illustrate the closeness of these two pdf's at large times, both embodying the 
long memory property of the aftershocks triggering process. One can also observe
the transition from the slope $-1-\theta$ at large times to
 the slope $-1+\theta$ at small times predicting by (\ref{efthas}). In contrast with Kagan and Knopoff who assumed
 that no daughters can be triggered between times $0$ and $c$ after the mother event occurred, or with Ogata
 who assumed an asymptotic constant rate of triggering at small times after the mother event occurred,
 the fractional exponential distribution amounts to assuming a diverging triggering rate as 
 one looks closer and closer to the main event. But, given the value of the exponent $-1+\theta >-1$,
 the total number of daughters triggered at short is finite (the pdf is integrable).

From the fractional exponential pdf, we define the survival function
\begin{equation}\label{seqintef}
\Phi_\theta(t) = \int_t^\infty f_\theta(t') dt' .
\end{equation}
Its Laplace image and explicit expression are given by
\begin{equation}\label{fphithexprsa}
\hat{\Phi}_\theta(u) = \frac{u^{\theta-1}}{1+u^\theta} \qquad \Leftrightarrow \qquad \Phi_\theta(t) = E_\theta \left(-t^\theta \right)~ ,
\end{equation}
where $E_\alpha(z)=E_{\alpha,1}(z)$ is the Mittag-Leffler function obtained as the special case $\beta=1$
of the generalized Mittag-Leffler function (\ref{etn4hw}). The following properties hold
\begin{equation}\label{phitheasumps}
\hat{\Phi}_\theta(u) \simeq
\begin{cases}
u^{\theta -1} , & u \ll 1 ,
\\
u^{-1} , & u \gg 1 .
\end{cases}
\quad \Leftrightarrow \quad \Phi_\theta (t) \simeq
\begin{cases} \displaystyle
1 - \frac{t^\theta}{\Gamma(1+\theta)} , & \tau \ll 1 ,
\\[4mm] \displaystyle
\frac{t^{-\theta}}{\Gamma(1-\theta)} , & t \gg 1 .
\end{cases}
\end{equation}

In the limiting case $\theta=1$, the fractional exponential distribution reduces to the pure exponential pdf:
\begin{equation}\label{expdisdef}
f_1(t) = \Phi_1(t) = e^{-t} \qquad \Leftrightarrow \qquad \hat{f}_1(u) = \hat{\Phi}_1(u) = \frac{1}{1+u} .
\end{equation}

Below, we will analyze in details the statistics of seismic clusters both for the cases when the pdf of waiting times
is described by the fractional exponential distribution $f_\theta(t)$ ($0<\theta<1$) and by the exponential case $\theta=1$ \eqref{expdisdef}.

\subsection{Statistical descriptions of the aftershocks that make up a seismic cluster}

\begin{figure}
\begin{center}
\includegraphics[width=0.95\linewidth]{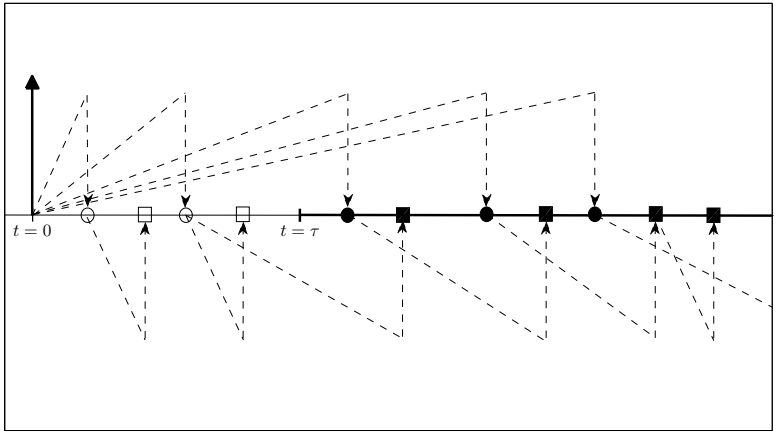}
\end{center}
\caption{Given a fixed current time $\tau$, we give an illustration of the definitions 
of past and future direct and indirect aftershocks triggered by
a main shock that occurred at time $0$, depicted by the bold vertical arrow at the origin of time.
The circles represent directly triggered events, i.e. first-generation daughters. 
The squares represent  triggered events of second and 
higher generations (grand-daughters, grand-grand-daughters, and so on).
Empty (resp. full) symbols correspond to events in the past, i.e. that occurred
at times $t<\tau$ (resp. in the future, i.e that occurred at times $t>\tau$).}\label{daughters}
\end{figure}

Consider a mother event (an ``immigrant'' in the language of branching processes) occurring at time $0$.
At the current time $\tau>0$, we distinguish the triggered events of first-generation (direct daughters)
represented as empty and full circles in figure \ref{daughters} and the triggered events of second and 
higher generations (grand-daughters, grand-grand-daughters, and so on) represented by empty
and full squares in figure \ref{daughters}. Moreover, we separate the events in the past (empty 
symbols) from the events that will occur in the future (full symbols).

Let us defined the GPF $\Omega(z;t)$ of the number $R(t)$ of the future events of all generations.
By Proposition 1, $\Omega(z;t)$ can be represented as a product of two GPFs:
\begin{equation}\label{Thetasplit}
\Omega(z;t) := \textnormal{E}\left[z^{R(t)}\right] = \Omega_\text{out}(z;t) \cdot \Omega_\text{in}(z;t) ~,
\end{equation}
where $\Omega_\text{out}(z;t)$ is the GPF of the number of future events triggered after the current instant $t$ 
by the past daughters of first-generation, and $\Omega_\text{in}(z;t)$ is the GPF of the total number of future daughters 
and all their higher-generation aftershocks. In other words, the seismic activity in the future (at times after
the present time $t$) can be decomposed as due to two sources: (i) the set of all the first-generation daughters 
that were born up to the current time $t$, which is represented by $\Omega_\text{out}(z;t)$;
(ii) the mother event that occurred at time $0$, which continues to trigger  
direct daughters and all their grand-daughters and higher generation events in the future,
and which is represented by $\Omega_\text{in}(z;t)$.

Using proposition~\ref{propositionone}, $\Omega_\text{out}(z;t)$ and $\Omega_\text{in}(z;t)$ 
are given by
\begin{equation}\label{thetoutinexpr}
\begin{array}{c} \displaystyle
\Omega_\text{out}(z;t) = \exp\left(-\kappa \mu \int_0^t \left[1 - G(z;t-s)\right] f(s) ds \right) ~,
\\[3mm] \displaystyle
\Omega_\text{in}(z;t) = \exp\left(-\kappa \mu \left[1-z G(z) \right] \int_t^\infty f(s) ds \right)~ ,
\end{array}
\end{equation}
where $G(z)=G(z;t=0)$, and $G(z;t)$ is the GPF of the number of aftershocks
of all generations that are triggered by an event of arbitrary magnitude that occurred at time $0$.

\begin{proposition}\label{propositiontwo}
\textnormal{
The GPF $G(z;t)$ is determined by the two coupled equations
\begin{equation}\label{geaheqs}
\begin{array}{c} \displaystyle
G(z;t) = Q\left[H(z;t) \right] ,
\\[1mm] \displaystyle
H(z;t) = 1 - z G(z) \Phi(t) - G(z;t) \otimes f(t)~ .
\end{array}
\end{equation}
where the symbol $\otimes$ represents the convolution operation, $\Phi(t)$ is the
survival function
\begin{equation}\label{seqintef2}
\Phi(t) = \int_\tau^\infty f(t') dt' 
\end{equation}
corresponding to the pdf $f(t)$ of waiting times, and the auxiliary function $Q(y)$ is defined by
\begin{equation}\label{qthrug}
Q(y) = G_d(1-y) ~.
\end{equation}
with $G_d$ defined by (\ref{Poisson}) and (\ref{thetoneexpr}) and given
approximately by (\ref{gonetwo}).
}
\end{proposition}
The proof of Proposition 2 is given in the Appendix.

It follows from the equations \eqref{geaheqs} that the GPF $G(z)$, of the total number of aftershock
of all generations triggered by some event, satisfies the well-known transcendent equation
\begin{equation}\label{transgxeq}
G(z) = G_d\left[z G(z) \right] ~.
\end{equation}

\begin{remark}
\textnormal{
For the feasibility of analytical calculations, we replace the exact expression \eqref{thetoneexpr} of the GPF $G_d(z)$ 
of the number of direct aftershocks by its algebraic expansion $\mathcal{G}_d(z)$ \eqref{gonetwo}.
After substitution in \eqref{qthrug}, we obtain
\begin{equation}\label{qydef}
Q(y) = \mathcal{G}_d(1-y) = 1 - n y +\rho y^\gamma~ .
\end{equation}
In the following, we will use this expression for the function $Q(y)$ in all our calculations,
offering when needed quantitative assessment of the quality of the approximation provided by this expansion.
}
\end{remark}

\begin{remark}
\textnormal{
To make explicit the dependence on the magnitude $m$ of the main shock and the
importance of the branching ratio $n$, in the following, we replace the notation
$H(z;t)$ in equation \eqref{geaheqs} by $H(z;t,n)$ and the function $\Omega(z;t)$ \eqref{Thetasplit} by 
$\Omega(z;t,n,m)$.
}
\end{remark}

\begin{proposition}\label{propositionthree}
\textnormal{
The GPF $\Omega(z;t,n,m)$ \eqref{Thetasplit} of the number of future offsprings of
all generations triggered by the mainshock of magnitude $m$ is given by
\begin{equation}\label{thetztevent}
\Omega(z;t,n,m) = e^{-\kappa \mu \cdot H\left[z G(z);t,n,m\right]} ,
\end{equation}
where $H(z;t,n)$ satisfies to the nonlinear integral equation
\begin{equation}\label{hcloseq}
H(z;t,n) - n\cdot H(z;t,n) \otimes f(t) +
\rho \cdot H^\gamma(z;t,n) \otimes f(t) = (1 - z) \Phi(t) .
\end{equation}
}
\end{proposition}
The proof of Proposition 3 is given in the Appendix.

For $t=0$, the following relation derives from \eqref{thetztevent} and \eqref{hcloseq}:
\begin{equation}\label{omegatzero}
\Omega(z;n,m):=\Omega(z;t=0,n,m) = e^{\kappa \mu \cdot \left(z G(z) - 1 \right)} ,
\end{equation}
where $G(z)$ is the solution of the transcendent equation \eqref{transgxeq}.

\subsection{Statistics of triggered events of magnitudes above a threshold}

As mentioned in section \ref{ethn3hg}, the fertility law (\ref{fertlaw}) holds
for all earthquakes with non-negative magnitudes. To account for the fact that
real seismicity is only detected above a magnitude threshold determined
by instrumental sensitivity and motivated by the fact that one may be interested
only in earthquakes of large magnitudes, we introduce the threshold magnitude $\nu>0$.
We can thus count the subset of triggered events whose magnitudes $\{m_j\}$ satisfy the inequality
$m_j > \nu > 0$. The magnitude $\nu$ can thus be considered to be an
observational magnitude threshold, such as only earthquakes with magnitudes larger than $\nu$
are observed. The existence of such a threshold can be shown to 
renormalise the parameters (branching ratio $n$ and background seismicity rate or immigrant rate)
of the ETAS model when applied to or calibrated on the observed earthquakes
\cite{SWapp05,SaiSorRenorm2006}.

\begin{figure}
\begin{center}
\includegraphics[width=0.95\linewidth]{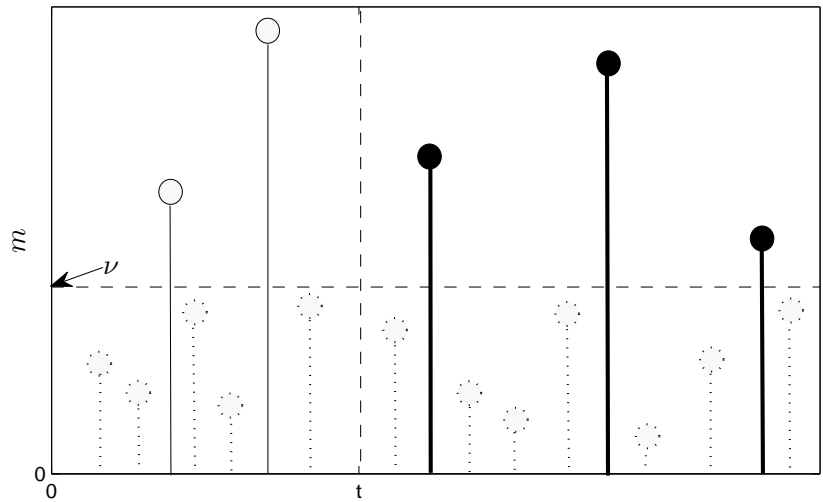}
\end{center}
\caption{Schematic representation of the offsprings of a given immigrant mother.
Along the vertical axis is the offsprings magnitudes. The horizontal dashed line corresponds to the threshold magnitude $\nu$.
The current time $t$ is indicated by the vertical dashed line. The dotted circles show the offsprings whose 
magnitudes are below $\nu$ and thus do not belong to the $\nu$-cluster. 
The hollow solid circles are the offsprings that belong to the $\nu$-cluster, but do not belong to the future $\nu$-cluster. 
The bold solid circles are the offsprings, which belong to the future $\nu$-cluster.}\label{circles}
\end{figure}

\begin{definition}
\textnormal{
We denote $\nu$-cluster the set of offsprings of all generations 
whose magnitudes exceed the given magnitude threshold  $\nu$. We call
future $\nu$-cluster the subset of the $\nu$-cluster of future offsprings, i.e. which
occur after the current time $t$. Figure~\ref{circles} illustrates the notion of the future $\nu$-cluster.
}
\end{definition}

\begin{proposition}\label{propositionfour}
\textnormal{
Given the GPF $\Omega(z;t,n,m)$ of the number of future offsprings of any positive
magnitude of a mother event 
of fixed magnitude $m$, the GPF $\Omega(z;t,n,m,\nu)$ of the number of future offsprings
of magnitude larger than $\nu$ is given by the following relation:
\begin{equation}\label{thetaemou}
\Omega(z;t,n,m,\nu) = \Omega\left(1+p(\nu) (z-1); t,n,m\right) .
\end{equation}
where
$p(m)$ is of the GR law \eqref{grldef}.
}
\end{proposition}
The proof of Proposition 4 is given in the Appendix.

\subsection{Probability of absence of offsprings, cluster durations, maximum magnitude
and function $\mathcal{P}(t,n,m,\nu)$}

Let us introduce the new function
\begin{equation}\label{xibigdef}
\mathcal{P}(t,n,m,\nu) := \Omega(z=0;t,n,m,\nu) = \Omega\left( 1-p(\nu);t,n,m\right) ~.
\end{equation}
Our motivation for proposing this function $\mathcal{P}(t,n,m,\nu)$ 
is that it has  three interesting probabilistic interpretations.

\subsubsection{First interpretation of $\mathcal{P}(t,n,m,\nu)$}

It follows from the first equality \eqref{xibigdef} and from the statistical meaning of the GPF $\Omega(z;t,n,m,\nu)$
that the function $\mathcal{P}(t,n,m,\nu)$ \eqref{xibigdef} is equal to the probability that the future $\nu$-cluster is empty,
i.e.  that the number $R_\nu(t)$ of future offsprings is equal to zero:
\begin{equation}
\mathcal{P}(t,n,m,\nu) = \text{Pr}\left\{R_\nu(t)=0 \right\} ~.
\end{equation}

\begin{figure}
\begin{center}
\includegraphics[width=0.95\linewidth]{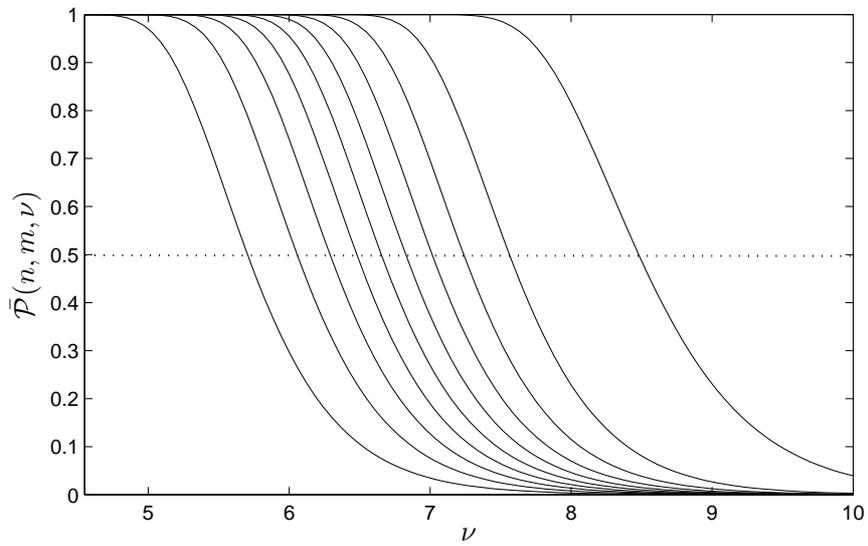}
\end{center}
\caption{Dependence of the probability $\overline{\mathcal{P}}(n,m,\nu)$ that the main shock triggers at least one offspring
of magnitude larger than $\nu$ as a function of $\nu$. For all curves, $m=9$, $b = 1$ and $\alpha =0.8$ ($\gamma=1.25$).
From right to left: $n=1, 0.9, 0.8, ..., 0.2, 0.1$. }\label{absplot}
\end{figure}

Defining $R_\nu := R_\nu(t=0)$, the function
\begin{equation}\label{probnuabs}
\mathcal{P}(n,m,\nu): = \text{Pr}\left\{R_\nu=0 \right\} =  \Omega(z;t=0,n,m,\nu) 
\end{equation}
is the probability that the mainshock of magnitude $m$ does not trigger offsprings of magnitudes larger than $\nu$.
Below, it will be convenient to define the complementary probability
\begin{equation}
\overline{\mathcal{P}}(n,m,\nu) : = 1 - \mathcal{P}(n,m,\nu)~,
\label{turjhe}
\end{equation}
that the mainshock triggers at least one observable offspring of magnitude larger than $\nu$.
Figure~\ref{absplot} shows the dependence of the probabilities $\overline{\mathcal{P}}(n,m,\nu)$ 
as a function of the threshold magnitude $\nu$, for different values of the 
branching ratio $n$, for a main shock magnitude equal to $m=9$.

\subsubsection{Second interpretation of $\mathcal{P}(t,n,m,\nu)$}

Let us introduce  $T(\nu)$ as the random duration of the $\nu$-cluster:
\begin{equation}
T(\nu) := \max\{T_{j}(\nu)\} ~,
\end{equation}
where $\{T_{j}(\nu)\}$ is the set of occurrence times of all the events that make up the $\nu$-cluster. 
Then, the second probabilistic meaning of the function $\mathcal{P}(t,n,m,\nu)$ is expressed by
\begin{equation}\label{medtprdef}
\mathcal{P}(t,n,m,\nu)  = \text{Pr}\left\{T(\nu)< t\right\} ~,
\end{equation}
i.e., it is the probability that the total duration of the $\nu$-cluster be less than $t$.
Note that this relation (\ref{medtprdef}) does not exclude the possibility that the main shock does not trigger any offspring. 

Considering only the $\nu$-clusters that contain at least one offspring, relation \eqref{medtprdef} 
has to be replaced by a conditional one, expressing the condition that the $\nu$-cluster contains at least one offspring. 
Using the law of total probability, the corresponding conditional counterpart of relation \eqref{medtprdef} reads
\begin{equation}\label{eftprcondef}
\text{Pr}\left\{T(\nu)< t|R_\nu>0 \right\} = \left[\mathcal{P}(t,n,m,\nu) - \mathcal{P}(n,m,\nu) \right] \big/ \overline{\mathcal{P}}(n,m,\nu) ~,
\end{equation}
where $R_\nu:=R_\nu(t=0)$ is the total number of offsprings in the $\nu$-cluster.

Denoting $\varphi(t; n,m)$ as the pdf of the total duration of the $\nu$-clusters that contain at least one offspring,
expression (\ref{eftprcondef}) yields
\begin{equation}\label{varphidef}
\varphi(t;n,m,\nu) = \mathcal{C} \cdot \frac{\partial \mathcal{P}(t,n,m,\nu)}{\partial t} , \qquad  \mathcal{C} := 1\big/\overline{\mathcal{P}}(n,m,\nu) ~.
\end{equation}

\subsubsection{Third interpretation of $\mathcal{P}(t,n,m,\nu)$}

Let us define $M(t,n,m)$ as the largest magnitude among all those of future offsprings triggered after 
time $t$ by a given main shock of magnitude $m$ that occurred at time $0$:
\begin{equation}\label{mtnm}
M(t,n,m) := \max\{m_j(t)\} ,
\end{equation}
where $\{m_j(t)\}$ is the set of the magnitudes of all future offsprings triggered after time $t$. 

Analogous to relation \eqref{eftprcondef}, we have
\begin{equation}\label{emmuconprob}
\text{Pr}\left\{M(t,n,m) < \nu|R(t) >0 \right\} = \frac{ \mathcal{P}(t,n,m,\nu) - \mathcal{P}(t,n,m)} {\overline{\mathcal{P}}(t,n,m)} ~,
\end{equation}
where $R(t)=R_{\nu=0}(t)$ is the total number of future offsprings,
$\mathcal{P}(t,n,m):= \text{Pr}\left\{R(t)=0\right\} = \Omega(z=0;t,n,m)$
and $\overline{\mathcal{P}}(t,n,m)$ defined by (\ref{turjhe})
is the probability that there is at least one triggered future event.

Let $\psi(\nu;t,n,m)$ be the pdf of the maximal magnitude over all future offsprings. 
It follows from relation \eqref{emmuconprob} that
\begin{equation}\label{psidef}
\psi(\nu;t,n,m) = \frac{1}{\overline{\mathcal{P}}(t,n,m)} \cdot \frac{\partial \mathcal{P}(t,n,m,\nu)}{\partial \nu}~ .
\end{equation}

\subsubsection{Properties of the second largest offspring}

Equalities \eqref{emmuconprob} and \eqref{psidef} are simple consequences of
the theory of order statistics applied to the offsprings magnitudes. 
Here, we provide some additional order statistics relations that are relevant to the 
understanding of the $\nu$-cluster's statistics. These relations
derive by using elementary facts of the theory of order statistics (see for instance Ref. \cite{Feller2}).

Let $M_1(t,n,m)$ be the second largest magnitude in the set of future offsprings.
By definition, it is smaller than the largest magnitude $M$
($M_1<M$), but is larger than the magnitudes of all other future offsprings. Then, 
the pdf $\psi_1(\nu;t,n,m)$ of the second largest magnitude $M_1(t,n,m)$ is given by
\begin{equation}\label{mfirstpdf}
\psi_1(\nu;t,n,m) = \frac{1}{2} \cdot \frac{d p^2(\nu)}{d\nu} \cdot \frac{ \mathcal{Q}(t,n,m,\nu)} {\overline{\mathcal{P}}_1(t,n,m)} ~,
\end{equation}
where 
\begin{equation}
\mathcal{Q}(t,n,m,\nu) = \frac{\partial^2 \Omega(z;t,n,m)}{\partial z^2} \bigg|_{z=1-p(\nu)}~ ,
\end{equation}
with the GPF $\Omega(z;t,n,m)$ of the total number of future offsprings number given by expression \eqref{thetztevent}. 
In addition, we have defined
\begin{equation}
\overline{\mathcal{P}}_1(t,n,m) := \text{Pr}\left\{R(t)>1\right\} = 1 - \mathcal{P}_1(t,n,m)
\end{equation}
as the probability that the number $R(t)$ of all future offsprings be larger than one. One can then show that
\begin{equation}
\mathcal{P}_1(t,n,m) = \mathcal{P}(t,n,m) + \frac{\partial \Omega(z;t,n,m)}{\partial z} \bigg|_{z=0}~ .
\end{equation}

There is another remarkable fact that derives from the structure of the GR law (\ref{grldef})
and the ETAS model, which can be called the  \emph{extremal GR law}. The random difference
\begin{equation}
\delta M = M(t,n,m) - M_1(t,n,m)
\end{equation}
between the largest and second-largest magnitudes is distributed according to the same
 regular GR law \eqref{grldef}, for any $n$ and $m$.

\section{Expression of function $H(z;t,n)$ defining the statistics of the number of future offsprings}

In this section, we provide an exact expression for the function $H(z;t,n)$ that obeys
the nonlinear integral equation \eqref{hcloseq}, and which determines the GPF $\Omega(z;t,n,m)$ \eqref{thetztevent} of the number of future offsprings, in the case of the
exponential pdf $f_1(t)$ \eqref{expdisdef}. Using the insights obtained from this exact solution
together with the properties of the function $H(z;t,n)$, we then 
formulate a conjecture for its general structure for an arbitrary pdf $f(t)$.

\subsection{Structure of the function $H(z;\tau,n)$}

Let us first consider the case where the pdf of waiting times is the 
exponential function $f_1(\tau)$ \eqref{expdisdef}. In this case,
the nonlinear integral equation \eqref{hcloseq} reduces to an initial value problem for the ordinary differential equation:
\begin{equation}\label{hodezet}
\begin{array}{c} \displaystyle
\frac{d H(z;t,n)}{dt} + (1-n) H(z;t,n) + \rho H^\gamma(z;t,n) = 0 ,
\\[2mm] \displaystyle
H(z;0,n) = 1- z .
\end{array}
\end{equation}
Its solution is
\begin{equation}\label{ahtrui}
H(z;t,n) = \mathcal{D}(z;\phi) \cdot
\mathcal{H}\left(z;\phi,\rho,n\right) ~,
\end{equation}
where
\begin{equation}\label{hztnlin}
\begin{array}{c} \displaystyle
\mathcal{D}(z;\phi)  := (1-z) \cdot \phi ,
\\[4mm] \displaystyle
\mathcal{H}(z;\phi,\rho,n) := \left({1 + \dfrac{\rho}{1-n} \cdot \left(1-\phi^{\gamma-1} \right) \cdot (1-z)^{\gamma-1}} \right)^\frac{1}{1-\gamma} ,
\end{array}
\end{equation}
and
\begin{equation}\label{phitauendef}
\phi := \Phi_1\left(t\big/t_1\right) = e^{(n-1) t } , \qquad t_1 :=1\big/(1-n) .
\end{equation}

Two important properties of the function $H(z;t,n)$ can be derived from this solution \eqref{ahtrui}.

\begin{property}\label{propone}
\textnormal{
The function $H(z;t,n)$ \eqref{ahtrui} is the product of two factors, which have a clear physical meaning. 
The function $\mathcal{D}(z;\phi)$ corresponds to the \emph{one-daughter approximation}, where each offspring triggers not more than one first-generation aftershock. In contrast, the function $\mathcal{H}(z;\phi,\rho,n)$ takes into account 
that each offspring can trigger more than one first-generation aftershock. Mathematically, this is
responsible for the factor $\rho$ in the power law asymptotics \eqref{qudelar} of the probability of 
first-generation aftershock numbers. If $\rho=0$, i.e. if each offspring triggers no more than one first-generation aftershock, then the second factor in the right-hand side of  \eqref{ahtrui} becomes $\mathcal{H}\equiv 1$ and thus 
$H(z;t,n)$ reduces to $\mathcal{D}(z;\phi)$.
}
\end{property}

\begin{property}\label{proptwo}
\textnormal{
Both functions $\mathcal{D}$ and $\mathcal{H}$ in expression \eqref{ahtrui} depend on the current time $t$ only via the function $\phi$ \eqref{phitauendef}.
}
\end{property}

\subsection{Conjecture for the structure of function $H(z;t,n)$}

Based on the physical meaning of the decomposition (\ref{ahtrui})
of the function $H(z;t,n)$ \eqref{ahtrui}, we propose the following overall structure of the function $H(z;t,n)$
for arbitrary waiting time distributions $f(t)$.

\begin{conjecture}\label{conjone}
\textnormal{
We suggest that the structure of the function $H(z;t,n)$ takes the form of a product
of $\mathcal{D}$ and $\mathcal{H}$ as given by expression (\ref{ahtrui}),
independently of the waiting time distribution $f(t)$, where $\mathcal{D}(z;\phi)$ corresponds to the \emph{one-daughter approximation} and $\mathcal{H}$ takes into account 
that each offspring can trigger more than one first-generation aftershock.
Thus, in order to get the general form of the function $H(z;t,n)$, one just needs to find the
generalisation of the time-dependent function 
$\phi = \Phi(t,n)$, contained in the functions $\mathcal{D}(z;\phi)$ and $\mathcal{H}$.
In practice, $\phi$ can be determined from the calculation of $\mathcal{D}$.
}
\end{conjecture}

From a mathematical point of view, the one-daughter approximation corresponds to neglecting the parameter $\rho$ 
in the nonlinear integral equation \eqref{hcloseq}, which amounts to linearise it.
The function $\phi=\Phi(t,n)$ can be obtained from
\begin{equation}\label{hclin}
H(z;t,n) - n\cdot H(z;t,n) \otimes f(t) = (1 - z) \cdot \Phi(t)~,
\end{equation}
which derives by linearising the nonlinear integral equation \eqref{hcloseq}.

To solve (\ref{hclin}), we apply the Laplace transform with respect to $t$ to 
equation \eqref{hclin}  term by term. The corresponding algebraic equation for the Laplace image
\begin{equation}
\hat{H}(z;u,n) := \int_0^\infty H(z;t,n) e^{-u t} dt
\end{equation}
reads
\begin{equation}\label{hateralgeq}
\hat{H}(z;u,n) - n \hat{H}(z;u,n) \hat{f}(u) = (1-z) \hat{\Phi}(u) ~,
\end{equation}
where $\hat{f}(u)$ is the Laplace image of the pdf $f(t)$, and $\hat{\Phi}(u)$ is the Laplace image of the corresponding survival function $\Phi(t)$ \eqref{seqintef}. Using relation \eqref{seqintef} and the standard properties of the Laplace transform,
we obtain 
\begin{equation}\label{hatphigen}
\hat{\Phi}(u) = \frac{1}{u} \left( 1- \hat{f}(u)\right) .
\end{equation}
Using \eqref{hateralgeq} and (\ref{hatphigen}) yields
\begin{equation}\label{haterexpr}
\hat{H}(z;u,n) = (1-z) \cdot \hat{\Phi}(u,n) , \qquad
\hat{\Phi}(u,n) :=\frac{1}{u}\frac{ 1- \hat{f}(u)}{1- n \hat{f}(u)} ~.
\end{equation}
Accordingly, the solution of equation \eqref{hclin} is described by the first equation of \eqref{hztnlin}, 
where now $\phi = \Phi(t,n)$, where $\Phi(t,n)$ is the inverse Laplace transform of the function $\hat{\Phi}(u,n)$ \eqref{haterexpr}:
\begin{equation}\label{invlaptransf}
\Phi(t;n) = \frac{1}{2 \pi i} \int_{-i\infty}^{i\infty}  e^{u t} \frac{ 1- \hat{f}(u)}{1- n \hat{f}(u)} du .
\end{equation}
According to our conjecture, the correct function $H(z;t,n)$ for an arbitrary pdf $f(t)$
is obtained by substituting the function $\phi=\Phi(t,n)$ \eqref{invlaptransf} into the
expressions \eqref{ahtrui}, \eqref{hztnlin}.

\begin{remark}
\textnormal{
We will show below that, in the subcritical case $n \lesssim 0.9$ and/or for 
sufficiently large threshold magnitudes $\nu\gtrsim 6$ (compared to the main shock
magnitude usually taken equal to $8$ or $9$ in our discussion), 
the statistics of the $\nu$-clusters is rather well described by the one-daughter approximation, i.e. by replacing the exact equality \eqref{ahtrui} by the approximation
\begin{equation}\label{odap}
H(z;t,n) \simeq \mathcal{D}(z;\phi) = (1-z) \cdot \Phi(t,n)~ .
\end{equation}
This may be interpreted by the fact that the one-daughter approximation gives accurate expressions for the $\nu$-cluster's statistics, not only for the exponential pdf $f_1(t)$, but also for arbitrary pdf $f(t)$.
}
\end{remark}

\subsection{Application to the fractional exponential case}

In the case where $f(t)$ is the fractional exponential distribution $f_\theta(t)$ \eqref{hateftheta},
we substitute its Laplace image $\hat{f}_\theta(u)$  \eqref{feqlim} in expression \eqref{haterexpr} to obtain
\begin{equation}
\hat{\Phi}_\theta(u,n) = \frac{u^{\theta-1}}{1-n + u^\theta}~ .
\end{equation}
Obviously, the inverse Laplace transform of $\hat{\Phi}_\theta(u,n)$ is equal to the 
fractional exponential survival function itself up to a rescaling of time:
\begin{equation}\label{phithet}
\phi = \Phi_\theta(t,n): = \Phi_\theta\left(t\big/t_\theta\right), \qquad t_\theta = (1-n)^{-1/\theta} ,
\end{equation}
where $\Phi_\theta(t)$ \eqref{fphithexprsa} is the fractional exponential survival function.

Using conjecture~\ref{conjone} and relations \eqref{phithet}, \eqref{ahtrui}, the
function $H_\theta(z;t,n)$ in the case where $f(t)$ is the fractional exponential distribution \eqref{hateftheta}
is obtained by replacing $\phi$ by $\Phi_\theta(t,n)$ in the right-hand side of expressions \eqref{ahtrui} and \eqref{hztnlin}:
\begin{equation}\label{ahtruithet}
H_\theta(z;t,n) = (1-z) \cdot \Phi_\theta(t,n) \cdot
\mathcal{H}\left[z;\Phi_\theta(t,n),\rho,n\right] .
\end{equation}
In the particular case $\theta=1$, where the fractional exponential survival function $\Phi_\theta(t)$ \eqref{fphithexprsa} reduces to the exponential one $\Phi_1(t)=e^{-t}$, expression \eqref{ahtruithet} becomes the exact solution of the initial value problem \eqref{hodezet}.

\section{Statistical properties of $\nu$-clusters}

In section~2, we have derived the general relations \eqref{varphidef}, \eqref{psidef} and \eqref{mfirstpdf} 
describing the statistics of the duration of $\nu$-clusters and of the magnitudes of the largest events in the $\nu$-cluster.
In section~3, we have obtained the explicit formulas needed to calculate the statistical characteristics
of $\nu$-clusters, and their dependence on the branching ratio $n$, magnitude $m$ 
of the main shock and observation magnitude threshold $\nu$. In the present section,
we exploit these formulas to present detailed results on the statistical properties of $\nu$-clusters.
We perform our analysis for the fractional exponential case, 
for which the waiting time distribution of triggered aftershocks is described by the
fractional exponential pdf $f_\theta(t)$ \eqref{hateftheta}, which 
is asymptotically equivalent (for large $t\gg1$) to the modified Omori-Utsu law \eqref{domoridist}.

\subsection{Properties of the intermediate function $\eta(t,n,\zeta)$}

Consider expression \eqref{xibigdef} for the probability $\mathcal{P}(t,n,m,\nu)$ of absence of future offsprings
(i.e. the probability that the future $\nu$-cluster is empty).
According to the relations \eqref{thetztevent}, \eqref{ahtrui}, \eqref{hztnlin}, we can write  $\mathcal{P}(t,n,m,\nu)$ as
\begin{equation}\label{omtexpr}
\mathcal{P}(t,n,m,\nu) = e^{-\chi \cdot \phi \cdot \eta(\phi,n,\zeta) } , \qquad \phi = \Phi_\theta(t,n) ,
\end{equation}
where
\begin{equation}\label{etadef}
\eta(\phi,n,\zeta) := \left( 1 + \mathcal{S} \cdot \left[1-\phi^{\gamma-1}  \right] \right)^{\frac{1}{1-\gamma}} ,
\end{equation}
and
\begin{equation}\label{mathesdef}
\begin{array}{c} \displaystyle
\chi := \kappa \cdot \mu \cdot \zeta ,  \qquad \mathcal{S} := \frac{\rho  \cdot \zeta^{\gamma-1} }{1-n} , \qquad \rho = \kappa^\gamma \gamma \Gamma(-\gamma) ,
\\[3mm] \displaystyle
\kappa = n(\gamma -1)\big/ \gamma ,\qquad  \zeta:=  1-(1-p) \cdot  G(1-p) .
\end{array}
\end{equation}

\begin{figure}
\begin{center}
\includegraphics[width=0.9\linewidth]{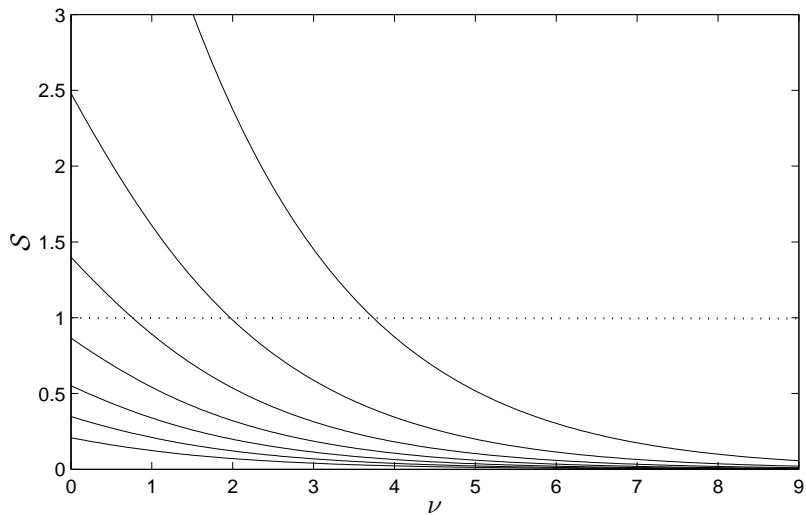}
\end{center}
\caption{Dependence of the factor $\mathcal{S}$ \eqref{mathesdef} as the function of the threshold 
magnitude $\nu$, for $(b = 1, \alpha = 0.8)$ ($\gamma=1.25$). From top to the bottom: $n=0.9, 0.8, 0.7, 0.6, 0.5, 0.4, 0.3$.
}\label{zetaplot}
\end{figure}

It is useful to study the properties of the function $\eta(t,n,\zeta)$ \eqref{etadef}, for our
subsequent analysis of the $\nu$-clusters statistics. The
dependence of $\eta(t,n,\zeta)$ as a function of the argument $\phi$
changes qualitatively depending on whether the factor $\mathcal{S}$ \eqref{mathesdef} is small or large.
Figure~\ref{zetaplot} shows the dependence of $\mathcal{S}$ as a function of the magnitude
threshold $\nu$. For large $\nu$ (most events cannot be observed), $\mathcal{S}$ becomes
smaller than $1$, while for small $\nu$ (most events are observable),  $\mathcal{S}$ 
is of the order of $1$ or larger. Large values of $\mathcal{S}$ are also obtained
near criticality, i.e. for $n \to 1$.

For $\mathcal{S}$ \eqref{mathesdef} small ($\mathcal{S}\ll 1$), then for any $\phi$, 
the function $\eta(\phi,n,\zeta)$, which quantifies the contribution
of the multiple generations of offsprings, is almost constant:
\begin{equation}\label{philin}
\mathcal{S} \ll 1 \quad \Rightarrow \quad \eta(\phi,n,\zeta) \simeq 1  \qquad \forall ~ \phi \in(0,1) ~
\end{equation}
which ensures the validity of the one-daughter approximation.

\begin{figure}
\begin{center}
\includegraphics[width=0.9\linewidth]{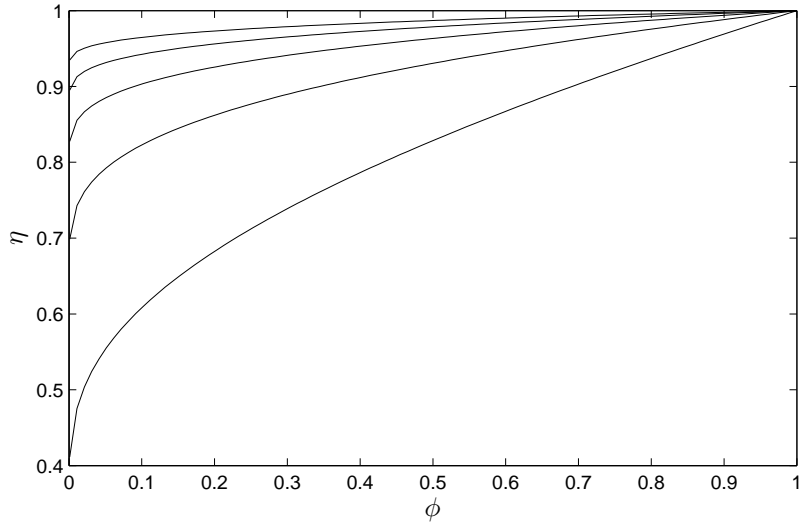}
\end{center}
\caption{Factor $\eta(\phi,n,\zeta)$ \eqref{etadef} as a function of $\phi$ for $\gamma=1.25$ and $\nu=6$. From top to the bottom: $n=0.5, 0.6, 0.7, 0.8, 0.9$. The corresponding values of $\mathcal{S}$ \eqref{mathesdef} are: $0.14; 0.23; 0.39; 0.72; 1.77$.  }\label{etaphiplot}
\end{figure}

In the opposite case $\mathcal{S}\gtrsim 1$, the dependence of $\eta(\phi,n,\zeta)$ \eqref{etadef} 
as a function of $\phi$ is strongly nonlinear.
Figure~\ref{etaphiplot} shows $\eta(\phi,n,\zeta)$ as a function of $\phi$ for several values of $n$, and thus 
$\mathcal{S}$. One can see that, for $n\lesssim 0.5$, the linear approximation \eqref{philin} holds for any 
threshold magnitude $\nu$.

Let us study separately the critical case $n=1$. Taking into account relations \eqref{mathesdef}, \eqref{phithet} and the asymptotics \eqref{phitheasumps} for the survival function $\Phi_\theta(t)$, one get:
\begin{equation}\label{limitslamb}
\begin{array}{c} \displaystyle
\lim_{n\to 1} \Phi_\theta(t,n) \equiv 1 , \qquad \lim_{n\to 1} \mathcal{S} \cdot \left[1-\Phi_\theta^{\gamma-1}(t,n)  \right] = \lambda \cdot t^\theta ,
\\[4mm] \displaystyle
\lambda := \frac{\rho \cdot \zeta^{\gamma-1} \cdot (\gamma-1)}{\Gamma(1+\theta)} .
\end{array}
\end{equation}
As a result, we obtain
\begin{equation}\label{etaenone}
\eta(t,\nu) :=  \lim_{n\to 1} \eta\left[\Phi_\theta(t,n),n,\zeta\right] = \left(1+ \lambda \cdot t^\theta \right)^{\frac{1}{1-\gamma}} .
\end{equation}
In the critical case ($n=1$), the function $\eta(t,\zeta)$ has thus the power law asymptotics
\begin{equation}\label{etapowasym}
\eta(t,\zeta) \simeq \lambda^{\frac{1}{1-\gamma}} \cdot t^{-\varrho} , \qquad \varrho := \frac{\theta}{\gamma-1} , \qquad \lambda \cdot t^\theta \gg 1 ~,
\end{equation}
with a tail exponent $\varrho$ that renormalises the waiting time distribution kernel via the exponent $\gamma$ quantifying the 
relative importance of different magnitude ranges in the generation of offsprings.

\subsection{Statistics of durations of $\nu$-clusters}

We are now armed to obtain the statistical distribution of the durations of future 
$\nu$-clusters, whose contributing events have magnitudes larger the threshold $\nu$. 
After substituting equalities \eqref{omtexpr}, \eqref{etadef} into relation \eqref{varphidef}, 
we obtain the following explicit expression for the pdf $\varphi_\theta(t;n,m,\nu)$ of the durations of $\nu$-clusters:
\begin{equation}\label{phithetexpr}
\begin{array}{c} \displaystyle
\varphi_\theta(t;n,m,\nu) =
\\[4mm] \displaystyle
\mathcal{C} \cdot \chi \cdot  \left(1 + \mathcal{S} \right) \cdot f_\theta(t,n) \cdot \eta^\gamma(\phi,n,\zeta) \cdot e^{-\chi \cdot \phi \cdot \eta(\phi,n,\zeta) } ~,
\end{array}
\end{equation}
where
\begin{equation}
\phi = \Phi_\theta(t,n) , \qquad
f_\theta(t,n) := \frac{d \Phi_\theta(t,n)}{d t} = \frac{1}{t_\theta} \, f_\theta\left(\frac{t}{t_\theta}\right) ~.
\end{equation}

\begin{figure}
\begin{center}
\includegraphics[width=0.95\linewidth]{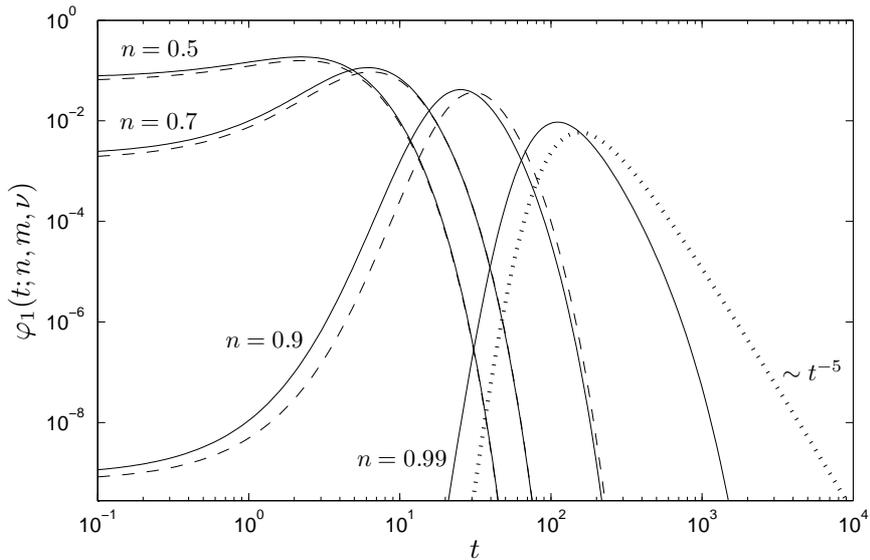}
\end{center}
\caption{Log-log plots of the exact pdf $\varphi_1(t;n,m,\nu)$ given by expression  \eqref{phithetexpr} of the durations of $\nu$-clusters in the pure exponential case $\theta=1\mapsto f(t)=e^{-t}$ for $m=9$, $\alpha=0.8$, $b=1$ ($\gamma=1.25$)á and $\nu=6$.
The different curves correspond to $n=0.99; 0.9; 0.7; 0.5$ respectively. The dotted line 
corresponds to the limiting pdf $\varphi_1(t;m,\nu)$ \eqref{phicrit}, obtained for the critical regime $n=1$. 
The dashed lines show the pdf $\varphi_1(t;n,m,\nu)$ \eqref{phithetone} in the one-daughter approximation, for 
the following values of the branching ratio: $n=0.9; 0.7; 0.5$. One can observe the validity of the 
one-daughter approximation. In the critical regime $n=1$, expression (\ref{etapowasym})
predicts a tail exponent $\varrho := \frac{\theta}{\gamma-1} = 4$ for $\theta=1$ and $\gamma=1.25$,
which explain the asymtotics $1/t^{1+\varrho} =1/t^5$ shown in the figure.
}\label{phinu3}
\end{figure}

Three limiting cases of the pdf \eqref{phithetexpr} of the $\nu$-cluster durations are worth discussing. 
\begin{enumerate}
\item {\bf One-daughter limit $\mathcal{S}\to 0$ and $\eta\to 1$}: then, expression \eqref{phithetexpr} reduces to
\begin{equation}\label{phithetone}
\varphi_\theta(t;n,m,\nu) = \mathcal{C} \cdot \chi \cdot f_\theta(t,n) \cdot e^{-\chi \cdot \Phi_\theta(t,n) } .
\end{equation}

\item {\bf Large time limit $t\to\infty$ for which $\phi \to 0$}: then, the asymptotics of the pdf $\varphi_\theta$ \eqref{phithetexpr} is defined by the asymptotics \eqref{efthas} of the original pdf $f_\theta(t)$ \eqref{hateftheta}. 
Remembering that, for $\theta=1$, the pdf $f_1(t)$ \eqref{expdisdef} reduces to the pure exponential, we have
\begin{equation}\label{phiasttinf}
\begin{array}{c}
\varphi_\theta(t;n,m,\nu) \simeq \mathcal{C} \cdot \left( 1 + \mathcal{S}  \right)^{\frac{1}{1-\gamma}} \cdot (1-n) \times
\\[4mm] \displaystyle
\times
\begin{cases} \displaystyle
-\frac{t^{-\theta-1}}{\Gamma(-\theta)} , &  \theta\in(0,1) ,
\\[4mm] \displaystyle
\quad e^{(n-1) t} ,  &  \theta =1 ,
\end{cases}
\qquad t \gg (1-n)^{-1/\theta} .
\end{array}
\end{equation}

\item {\bf Critical case $n=1$}: In this case, using the second limit in \eqref{limitslamb}, we obtain the following 
expression for the pdf of the $\nu$-cluster's durations:
\begin{equation}\label{phicrit}
\varphi_\theta(t;m,\nu) = \mathcal{C} \cdot \chi \cdot \frac{ \theta}{(\gamma-1) \cdot t} \cdot \frac{\lambda \cdot t^{\theta}}{\left(1+\lambda \cdot t^\theta\right)^{\frac{\gamma}{\gamma-1}}} \cdot e^{-\chi \cdot \left(1+ \lambda \cdot t^\theta \right)^{\frac{1}{1-\gamma}}} ~,
\end{equation}
which has the following power law asymptotics
\begin{equation}\label{phicritas}
\varphi_\theta(t;m,\nu) \simeq \xi \cdot t^{-\varrho-1}  , \qquad \xi := \mathcal{C} \cdot \chi \cdot \varrho \cdot \lambda^{\frac{1}{1-\gamma}} , \qquad t \to \infty ,
\\[4mm] \displaystyle
\end{equation}
where the exponent $\varrho$ is defined in \eqref{etapowasym}.
\end{enumerate}

\begin{figure}
\begin{center}
\includegraphics[width=0.95\linewidth]{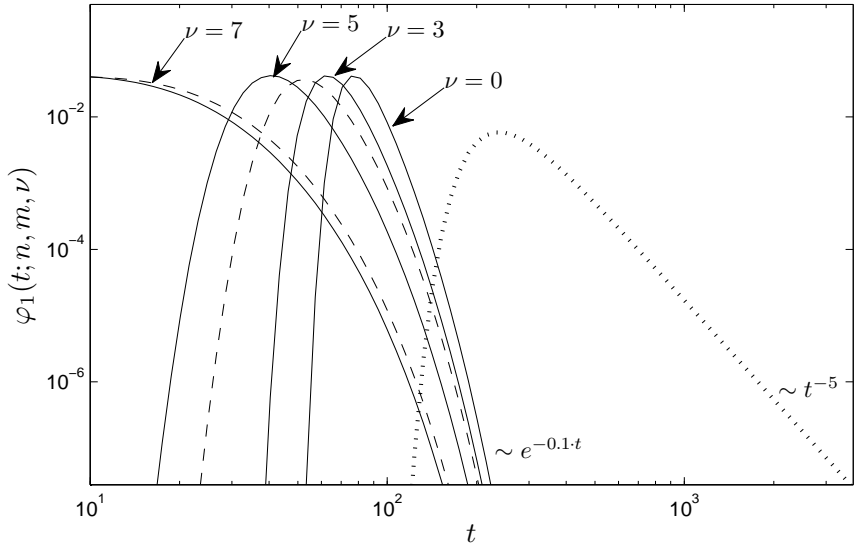}
\end{center}
\caption{Solid lines: plots of the pdf $\varphi_1(t;n,m,\nu)$ of the durations of $\nu$-clusters in the pure exponential case $\theta=1$, for $m=9$, $\alpha=0.8$, $b=1$ ($\gamma=1.25$),  for $n=0.9$ and for different $\nu=0; 3; 5; 7$. 
The dotted line  corresponds to the limiting pdf $\varphi_1(t;m,\nu)$ \eqref{phicrit}, obtained for the critical regime $n=1$
and $\nu=0$. The dashed lines are the one-daughter approximations for the pdf of the durations
of the $\nu$-clusters, for $\theta=1$, $n=0.9$ and $\nu=5; 7$.}\label{phithonenus}
\end{figure}

The following figures illustrate how the pdf $\varphi_\theta(t;n,m,\nu)$ \eqref{phithetexpr}
changes its shape upon variations of the main shock magnitude $m$, magnitude threshold $\nu$, 
branching ratio $n$. The figures also provide a check on the validity of the above asymptotic relations \eqref{phithetone}--\eqref{phicritas}.

Figure~\ref{phinu3} shows $\varphi_1(t;n,m,\nu)$ as a function of duration $t$
for a great main shock ($m=9$), a significant magnitude threshold($\nu=6$) for the pure exponential case ($\theta=1$)
and several values of the branching ratio.

Figure~\ref{phithonenus} shows $\varphi_1(t;n,m,\nu)$ as a function of duration $t$ in the pure exponential case $\theta=1\mapsto f(t)=e^{-t}$ for $n=0.9$ and for different magnitude thresholds $\nu$. Note that, for all $n<1$, $\varphi_1(t;n,m,\nu)$
tends to zero exponentially fast at large time $t\gg1$. One can observe that $\nu$ substantially influences 
the shape of $\varphi_1(t;n,m,\nu)$ only at small times $t\lesssim 1$. 
For large enough $\nu$'s, the exact pdf of the durations of the $\nu$-cluster's approaches closely
at all times $t$ the corresponding one-daughter limit pdf \eqref{phithetone}.

\begin{figure}
\begin{center}
\includegraphics[width=0.95\linewidth]{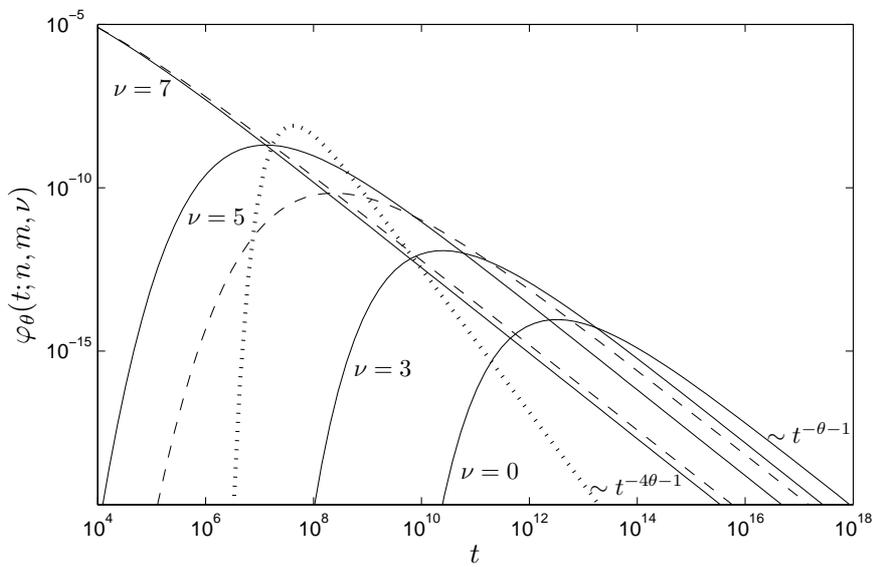}
\end{center}
\caption{Solid lines: plots of the pdf  $\varphi_\theta(t;n,m,\nu)$ of the durations of $\nu$-clusters  in the fractional exponential case $\theta=0.3$, for $m=9$, $\alpha=0.8$, $b=1$ ($\gamma=1.25$), for $n=0.9$ and for $\nu=0; 3; 5; 7$. The dotted line is the 
pdf  $\varphi_\theta(t;n,m,\nu)$ of the durations of $\nu$-clusters in the critical case $n=1$ and for $\nu=0$. The dashed lines 
show the pdf  $\varphi_\theta(t;n,m,\nu)$ of the durations of $\nu$-clusters in the one-daughter limit for $\theta=0.3$, $n=0.9$ and $\nu=5;7$.}\label{phith03nus}
\end{figure}

Figure~\ref{phith03nus} shows the pdf $\varphi_\theta(t;n,m,\nu)$ of the durations of $\nu$-clusters 
in the fractional exponential case $\theta=0.3$. One can observe that the key properties and shapes of 
$\varphi_\theta(t;n,m,\nu)$ differ dramatically from those of $\varphi_1(t;n,m,\nu)$.
In the later exponential case $\theta=1$, and for $n<1$, $\nu>0$, $\varphi_1$ tends to zero exponentially fast for $t\gg 1$.
It is only in the critical case $n=1$, that $\varphi_1(t;n,m,\nu)$ develops a power law tail, albeit with a
rather large exponent $\varrho$ (\ref{etapowasym}). In the fractional exponential case ($\theta<1$), 
it is remarkable that
the duration dependence of $\varphi_\theta$ is significantly slower than in the critical case $n=1$ at large $t$. 
For $n<1$ and for any $\nu\geqslant 0$, $\varphi_\theta$ tends to zero, at $t\gg 1$, following
the ``one-daughter law'' $\varphi_\theta\sim t^{-\theta-1}$ \eqref{phiasttinf}, which decays
to zero much more slowly than the
dependence in the critical regime given by $\varphi_\theta \sim t^{-\varrho-1}$ (where $\varrho>\theta$ is
given by (\ref{etapowasym})).

\begin{remark}\label{remfive}
\textnormal{
The pdf $\varphi_\theta(t;n,m,\nu)$ of the durations of $\nu$-clusters for the  fractional exponential
and the pdf $\varphi_1(t;n,m,\nu)$ for the exponential case, both determined from expression \eqref{phithetexpr},
share one important property.
In the subcritical case (for $n\lesssim 0.9$) and for sufficiently large magnitude thresholds $\nu$ (for figures~\ref{phinu3},~\ref{phithonenus} and~\ref{phith03nus}, for $\nu\gtrsim 6$), 
the two pdf's $\varphi_\theta(t;n,m,\nu)$ and $\varphi_1(t;n,m,\nu)$
\eqref{phithetexpr} are both very well approximated by their corresponding one-daughter limit \eqref{phithetone}.
Since the solution\footnote{It is equal to the inverse Laplace transform, with respect to argument $u$, of the expression \eqref{haterexpr}} of the integral equation \eqref{hcloseq} in the one-daughter limit $(\rho=0$) is exact, we
conjecture that this provides the almost exact expression for the pdf of the durations of $\nu$-clusters for all $\theta\in(0,1]$
in these regimes $n\lesssim 0.9$ and $\nu\gtrsim 6$.
}
\end{remark}

\subsection{Statistics of the maximum magnitude in $\nu$-clusters}

In this section, we study in detail the statistics of the maximal magnitude of future offsprings
of a main shock of fixed magnitude $m$ that occurred at time $0$. The pdf 
$\psi(\nu;t,n,m)$ of the maximal magnitude $\nu$ is given by  expression \eqref{psidef}. 
As a result of the equalities \eqref{omtexpr}-\eqref{mathesdef}, $\psi(\nu;t,n,m)$ 
depends on time $t$ only through the function $\Phi_\theta(t,n)$.
For definiteness, we take the function $\Phi(t,n)$ to be the fractional exponential, which
 includes as a special case the exponential function: $\theta=1 \mapsto \Phi(t,n) = e^{(n-1) t}$.
 Thus, for convenience, we rename $\psi$ as the function of the argument $\phi$: $\psi = \psi(\nu;\phi,n,m)$ and will discuss its time dependence via the auxiliary argument $\phi$. If one wish to recover the explicit time dependence of the pdf
 of the maximum magnitude of future offsprings, one just has to solve the equation $\phi = \Phi_\theta(t,n)$ for the time $t$. 
In the pure exponential case $\theta=1$, this correspondence has a simple explicit form $t = \ln(\phi)\big/ (n-1)$,
which maps the unit interval $\phi\in[0,1]$ onto the time axis $t\in(0,\infty)$.

Using relations  \eqref{psidef}, \eqref{omtexpr}-\eqref{mathesdef}, we obtain the explicit expression
\begin{equation}\label{psiexpr}
\psi(\nu;\phi,n,m) = \frac{\kappa \mu \phi p'(\nu)}{e^{-\kappa \mu \phi \eta(\phi,n)}-1} \cdot \zeta'\left[p(\nu)\right]
\cdot
\eta^\gamma (\phi,n,\zeta) \cdot e^{-\kappa  \mu \cdot \zeta \cdot \phi \cdot \eta(\phi,n,\zeta) }  ~,
\end{equation}
where
\begin{equation}
\begin{array}{c} \displaystyle
\eta(\phi,n):= \eta[\phi,n,1) = \left[1+\frac{\rho}{1-n} \left(1-\phi^{\gamma-1}\right)\right]^{\frac{1}{1-\gamma}} , \\[4mm] \displaystyle
\zeta'(p) :=\frac{d \zeta}{d p} , \qquad p'(\nu) := \frac{dp(\nu)}{d\nu} , \qquad p = 10^{-b \nu} ~,
\end{array}
\end{equation}
and $\zeta=\zeta(p)$ is defined in \eqref{mathesdef}.

Below, we will compare the exact expression \eqref{psiexpr} for $\psi(\nu;\phi,n,m)$ with its one-daughter approximation
\begin{equation}\label{psiexprlin}
\psi_{1 {\rm daughter}}(\nu;\phi,n,m) = \frac{\kappa \mu \phi}{e^{-\kappa \mu \phi }-1} \cdot \zeta'\left[p(\nu)\right] \cdot p'(\nu)\cdot e^{-\kappa  \mu \cdot \zeta \cdot \phi \cdot} ,
\end{equation}
where $\zeta$ and $\zeta'$ are given by the following relations:
\begin{equation}
\zeta(p) = \frac{p}{1-n (1-p)} , \qquad \zeta'(p) = \frac{1-n}{(1- n (1-p))^2} ~.
\end{equation}

\begin{figure}
\begin{center}
\includegraphics[width=0.9\linewidth]{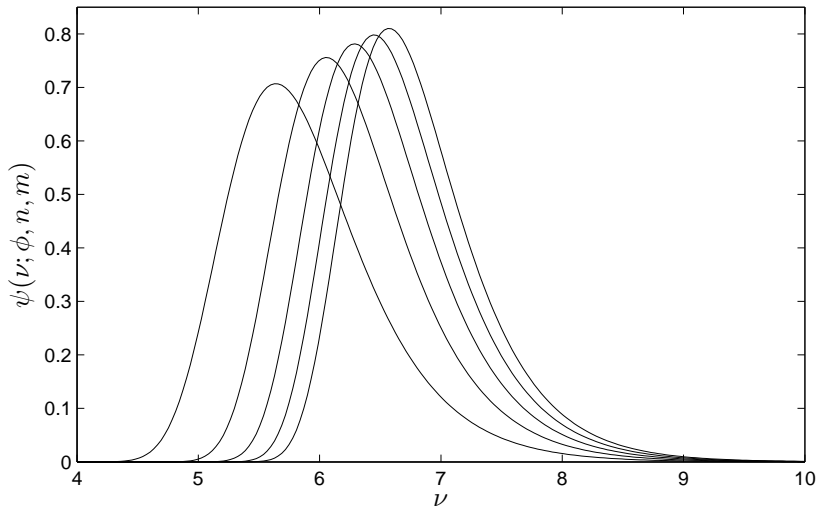}
\end{center}
\caption{Plots of the pdf $\psi(\nu;\phi,n,m)$ \eqref{psiexpr} of the maximal magnitude of future offsprings for $m=8$, $\gamma=1.25$ and for $n=0.9$. From left to right, the effective time is $\phi=0.2, 0.4, 0.6, 0.8, 1$, which corresponds to a
real time increasing from right to left.}\label{psiphi}
\end{figure}

Figure \ref{psiphi} shows the pdf $\psi(\nu;\phi,n,m)$ \eqref{psiexpr} of the maximal magnitude of future offspring for different values of the effective time $\phi$. As time increases, the pdf shifts to the left, indicating a decrease of the typical 
magnitude of the largest future offspring.

\begin{figure}
\begin{center}
\includegraphics[width=0.95\linewidth]{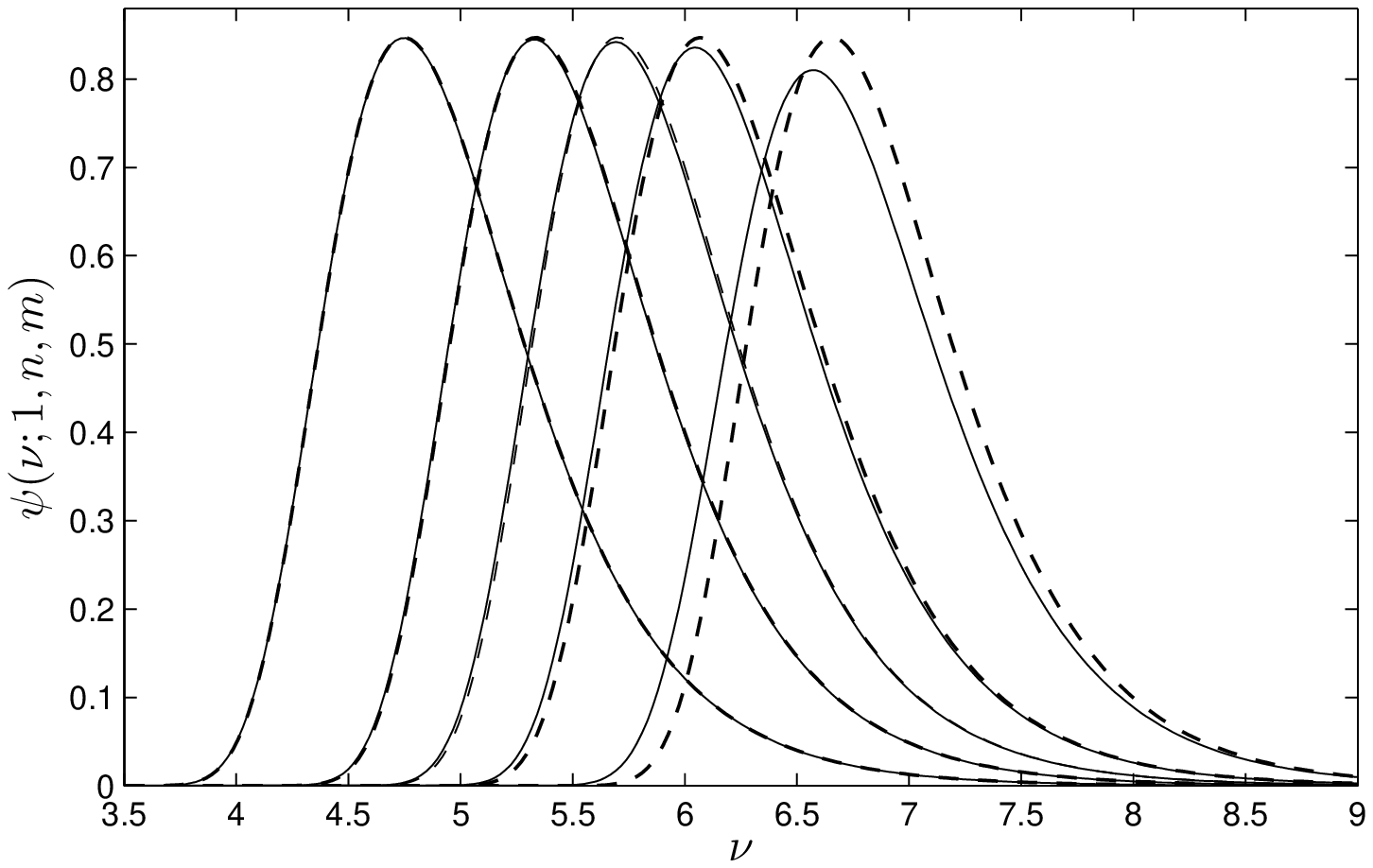}
\end{center}
\caption{Plots of the pdf $\psi(\nu;\phi=1,n,m)$ of the maximal magnitude over all offsprings triggered by a main shock
of magnitude $m=8$, with  $\alpha=0.8$, $b=1$ ($\gamma=1.25$). Solid lines: plots of the exact pdf \eqref{psiexpr}. Dashed lines: plots of the one-daughter approximation \eqref{psiexprlin}. From right to left: $n=0.9, 0.7, 0.7. 0.5. 0.3, 0.1$.}\label{psimaxlin}
\end{figure}

Figure~\ref{psimaxlin} compares the exact expression  \eqref{psiexpr} 
with its one-daughter approximation \eqref{psiexprlin} of the pdf $\psi(\nu;n,m) := \psi(\nu;\phi=1,n,m)$ at the effective ``time'' $\phi=1$
as a function of the maximal magnitude $\nu$ among all offsprings triggered by the mainshock
for different values of the branching ratio $n$. One can observe that the 
one-daughter expression \eqref{psiexprlin} provides a good approximation to the exact expression  \eqref{psiexpr} ,
the better the approximation, the smaller the branching ratio.

\begin{figure}
\begin{center}
\includegraphics[width=0.95\linewidth]{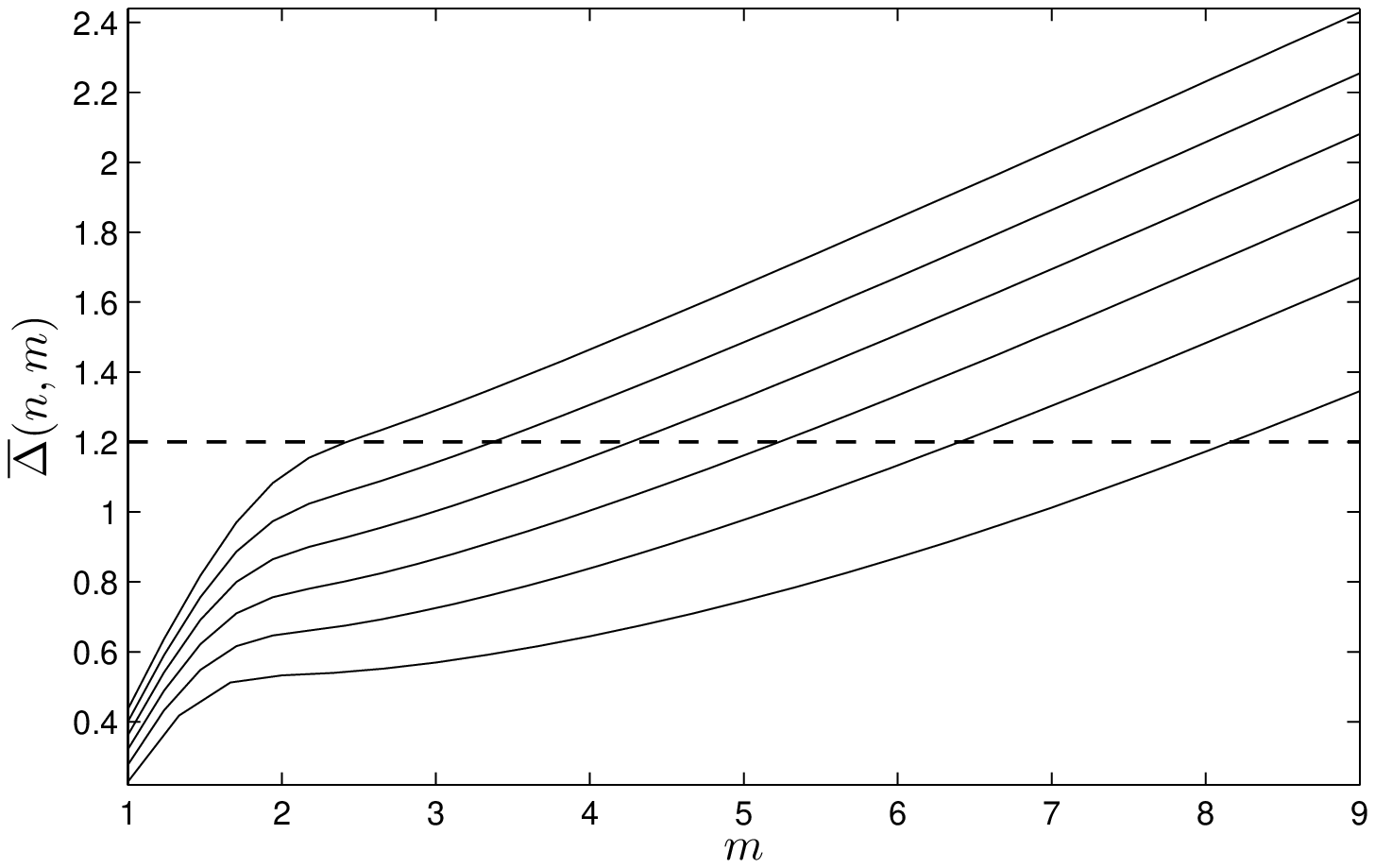}
\end{center}
\caption{Plot of the difference $\overline{\Delta}(n,m)$ \eqref{benem} between the main shock magnitude $m$ and the average magnitude $\overline{M}(n,m)$  of its largest offspring
as a function of the mainshock  magnitude $m$, for $\alpha=0.8$, $b=1$ ($\gamma=1.25$). From top to bottom: $n=0.4, 0.5. 0.6, 0.7, 0.8, 0.9$. The dashed straight line corresponds to the B\r{a}th law: $\overline{\Delta}(n,m) \simeq 1.2$.}\label{bnmplots}
\end{figure}

Let  $M(t,n,m)$ be the maximal magnitude of all future offsprings triggered by a main shock of magnitude $m$ that
occurred at time $0$, as defined by \eqref{mtnm}. Its mean value at the current time $t=t(\phi)$ reads
\begin{equation}
\overline{M}(\phi,n,m) := \int_0^\infty \nu \psi(\nu;\phi,n,m) d\nu~.
\end{equation}
At the specific time $t_1=t(\phi=1)$, this reduces to 
\begin{equation}
\overline{M}(n,m) := \overline{M}(\phi=1,n,m) = \int_0^\infty \nu \psi(\nu;n,m) d\nu ~.
\end{equation}
We use this expression to construct figure~\ref{bnmplots}, which shows the 
difference between the main shock magnitude $m$ and the average magnitude $\overline{M}(n,m)$  of its largest offspring:
\begin{equation}\label{benem}
\overline{\Delta}(n,m) := m - \overline{M}(n,m)
\end{equation}
for different values of the branching ratio $n$. One can observe that $\overline{\Delta}(n,m)$
monotonically increases with the main shock magnitude $m$. The dependence as a function of $m$ and
$n$ is sufficiently slow and smooth that the so-called B\r{a}th law, represented in figure~\ref{bnmplots} by the dashed straight line, 
if not correct, provides a rough estimate of $\overline{\Delta}(n,m)$ \cite{HSorGRLBath2003,SaiSorBath2005}.

\begin{figure}
\begin{center}
\includegraphics[width=0.95\linewidth]{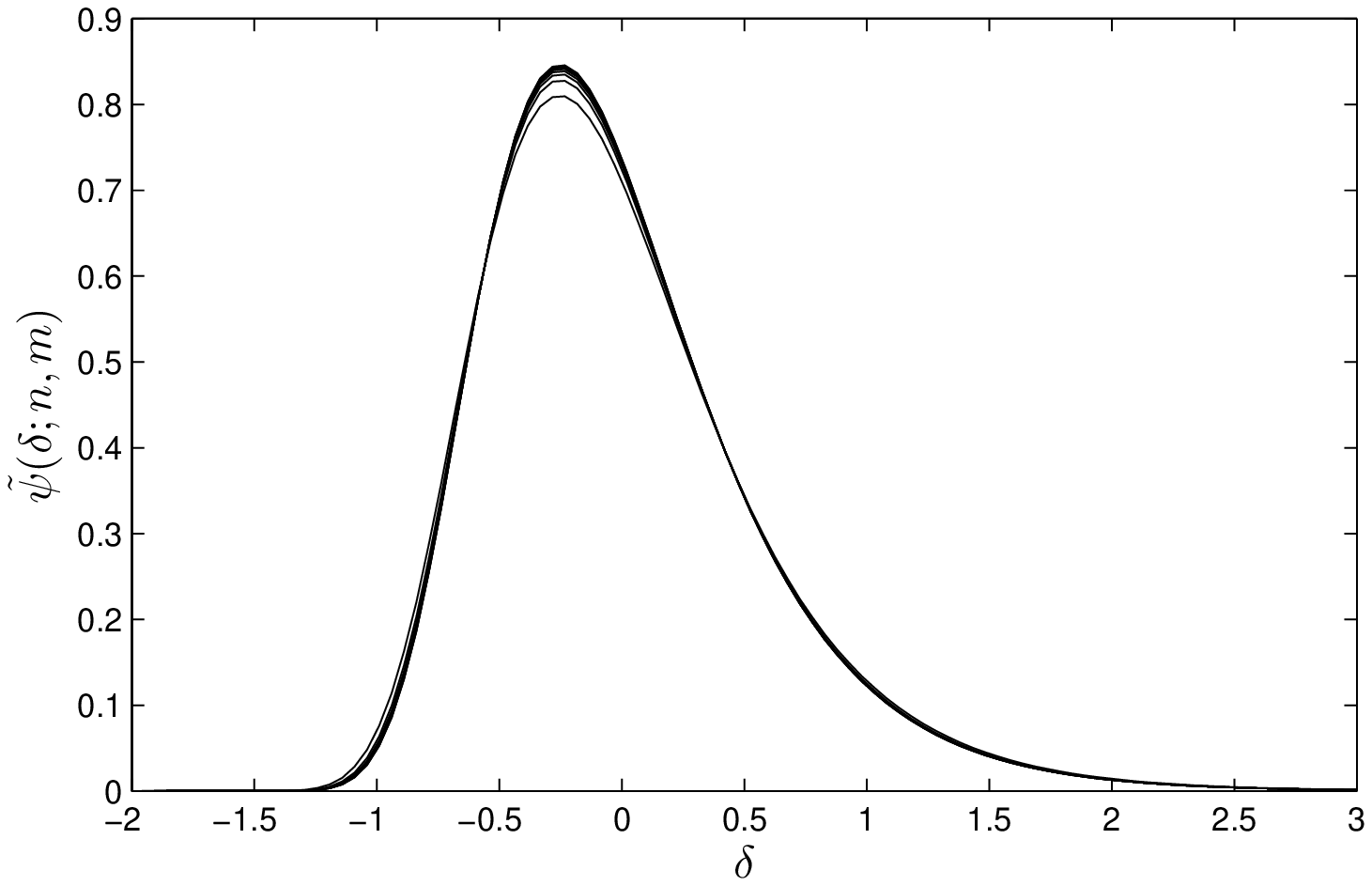}
\end{center}
\caption{Plots of the pdf $\tilde{\psi}(\delta;n,m)$ \eqref{psitilpdf} of the distance from the mean defined by expression
 \eqref{dedef} of the maximum magnitude of the future offprings of a main shock of magnitude $m=8$, for $\gamma=1.25$ and $n=0.1, 0.3, 0.5, 0.7, 0.9$.}\label{tildapsi}
\end{figure}

Another property of interest is that
the shapes of the pdf's $\psi$ \eqref{psiexpr} are almost identical for a wide range of main shock magnitude $m$ and 
values of the branching ratio $n$. By centering the pdf's according to
\begin{equation}\label{psitilpdf}
\tilde{\psi}(\delta;n,m) := \psi\left(\delta+\overline{M};n,m\right)
\end{equation}
where the distance from the mean is 
\begin{equation}\label{dedef}
\delta =M(n,m)-\overline{M}(n,m) ~,
\end{equation}
figure~\ref{tildapsi} shows an almost perfect collapse for the different values of the branching ratio $n$.

\begin{figure}
\begin{center}
\includegraphics[width=0.9\linewidth]{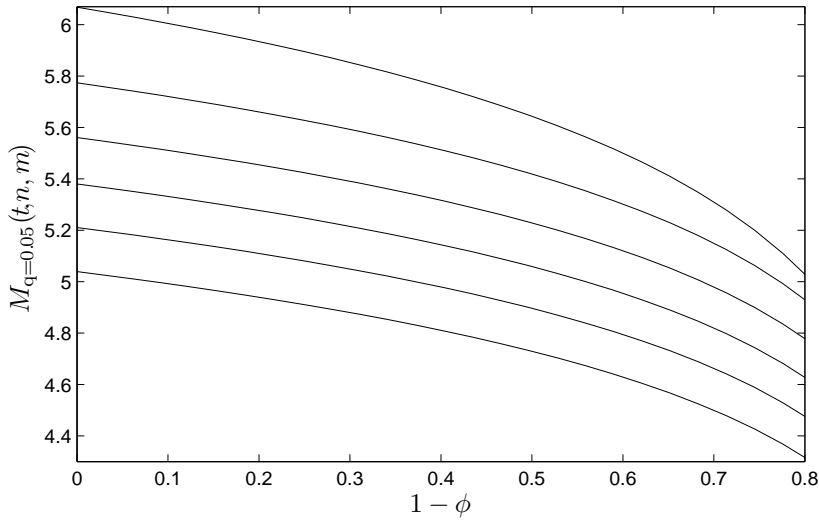}
\end{center}
\caption{Plots of the low quantile maximum magnitude $M_\text{q=0.05}(t,n,m)$ as a function of $1-\phi$, for $m=8$ and $\gamma=1.25$. From top to bottom, $n=0.9, 0.8, 0.7,0.6, 0.5, 0.4$. The curves shows that $M_\text{q=0.05}(t,n,m)$  decreases with increasing of time (that is, with decreasing $\phi$).}\label{minmaxplot}
\end{figure}

In figures~\ref{psimaxlin} and~\ref{tildapsi}, one can observe that the pdf's exhibit
rather sharp decay to the left, so that one can define a low quantile maximum magnitude $M_\text{q=0.05}(t,n,m)$ of 
the maximum magnitude of the offsprings in a $\nu$-cluster, such that 
\begin{equation}
\text{Pr}\left\{M(t,n,m)>M_\text{q=0.05}(t,n,m)\right\} = 0.95 ~,
\label{rgnetuju}
\end{equation}
where $M(t,n,m)$ is defined by equality \eqref{mtnm}. This definition (\ref{rgnetuju})
means that, typically in only one in twenty clusters, the maximum magnitude among all offsprings is smaller than 
$M_\text{q=0.05}(t,n,m)$. Figure~\ref{minmaxplot} shows the dependence of 
$M_\text{q=0.05}(t,n,m)$ as a function of the effective increasing time $1-\phi$ (since $\phi$ is a decreasing function of time).
$M_\text{q=0.05}(t,n,m)$ is a decreasing function of time, as the triggering activity decays progressively.

\section{Conclusion}

Using the standard ETAS model of triggered seismicity, we have presented 
a rigorous theoretical analysis of the main statistical properties of
temporal clusters, defined as the group of events triggered by a given main shock of fixed
magnitude $m$ that occurred at the origin of time, at times larger than some present time $t$.
The most general and powerful tool to derive analytically rigorously the statistical
properties of the numbers of events triggered by some main shock as a function of time
is the technology of generating probability function (GPF), of which we have recalled
the main properties and that we have applied to our problem.
We have derived the explicit and approximate expressions for the GPF of the number of future offsprings 
in a given temporal seismic cluster,  defining, in particular, the statistics of the cluster's duration 
and the cluster's offsprings maximal magnitudes. Our main results have been presented
in the form of four propositions, whose proofs have been given in appendices.

We introduced the probability $\mathcal{P}(t,n,m,\nu)$ that the future cluster of events of magnitudes above
some detection threshold $\nu$ is empty. This probability becomes the workhorse
for the derivation of our main results. $\mathcal{P}(t,n,m,\nu)$ can also be interpreted 
in its time dependence as the probability
that the total duration of the cluster of triggered events is less than $t$.
A third interesting interpretation relates the derivative of $\mathcal{P}(t,n,m,\nu)$
with respect to $\nu$ to the probability density function of the maximal magnitude
over all events within the temporal cluster. 
We used $\mathcal{P}(t,n,m,\nu)$ to derive the remarkable result that the
magnitude difference between the largest and second largest event in the
future temporal cluster is distributed according to the regular Gutenberg-Richer law
that controls the unconditional distribution of earthquake magnitudes.
 
The distribution $\varphi_\theta(t;n,m,\nu)$ of the durations of temporal clusters of events of 
magnitudes above some detection threshold $\nu$ was obtained in exact
analytical form, and investigated in three limits: (i) the one-daughter limit for $n<1$
in which each event can trigger not more than one first-generation aftershock,
(ii) the large time regime and (iii) the critical case when the branching ratio $n$ is equal
to its critical value $1$. For earthquakes obeying the Omori-Utsu law
for the distribution of waiting times between triggering and triggered events, 
we show that $\varphi_\theta(t;n,m,\nu)$ has a power law tail that is fatter
in the non-critical regime $n<1$ than in the critical case $n=1$. This paradoxical
behavior is similar to the one explained in Ref.\cite{SaiSor2010}, and results
from the fact that generations of all orders cascade very fast
in the critical regime and accelerate the temporal decay of the cluster dynamics.
We also derive the detailed shape of the distribution of the maximal magnitude
over all events in the future cluster triggered by some main shock.
We show that the so-called B\r{a}th law, 
stating that the difference between the main shock magnitude and the average
magnitude of its largest offspring is equal to $1.2$, is only roughly relevant,
given the fact that we document a non-trivial main shock magnitude dependence
as well as the influence of the branching ratio.

\vskip 1cm
{\bf Acknowledgements}: The second author dedicates this article to the first author, 
who was his cherished friend and long-time collaborator. AS was extraordinary young
in his mind, exceptionally creative and with a freshness and enthousiasm for research
rarely found even in beginning scientists. This article was almost finalised before
the time when AS left us and DS vowed to bring it to completion and 
have it published to honor his memory. DS regrets
that many other commitments has delayed this important endeavor.

\appendix

\section{Proofs of propositions 1-4}

\subsection{Proof of Proposition~\ref{propositionone}}

Let $\{T_k\}$ be the occurrence times of the triggered tdaughters, while $R_d(m)$ is the total number of daughters. Let us introduce the number $R_d([t_1,t_2];m)$ of daughters triggered within a time interval
\[
[t_1,t_2] , \qquad \forall ~ (t_1, t_2) : \quad 0< t_1 < t_2 < \infty .
\]
One may represent $R_d([t_1,t_2];m)$ in the form:
\begin{equation}
R_d([t_1,t_2];m) = \sum_{k=1}^{R_d(m)} \Pi\left(T_k, [t_1,t_2]\right) ,
\end{equation}
where
\[
\Pi\left(t , [t_1,t_2]\right) :=
\begin{cases}
1 , & t \in [t_1,t_2] ,
\\
0 , & t \notin [t_1,t_2] ,
\end{cases}
\]
is the indicator of the time interval $[t_1,t_2]$.

Consider the GPF of the random number $R_d([t_1,t_2];m)$
\begin{equation}\label{gddef}
G_d(z;[t_1,t_2]) := \textnormal{E}\left[ z^{R_d([t_1,t_2];m)} \right] .
\end{equation}
Taking into account the identity
\begin{equation}\label{zetpopi}
z^{\Pi\left(t , [t_1,t_2]\right)} \equiv 1 + (z-1) \Pi\left(t , [t_1,t_2]\right)
\end{equation}
and keeping in mind that $\{T_k\}$ are iid random variables, we can rewrite the GPF \eqref{gddef} in the form:
\begin{equation}\label{gedexprgen}
G_d(z;[t_1,t_2]) = \mathbb{E}\left[\left(1+(z-1) \mathbb{E}\left[\Pi\right]\right)^{R_d(m)}\right]~.
\end{equation}
We have used here the short notation $\Pi = \Pi\left(T , [t_1,t_2]\right)$. 
The outer expectation $\mathbb{E}\left[\cdots\right]$ at \eqref{gedexprgen} represents the statistical average over 
the total number $R_d(m)$ of daughters total number. The inner expectation $\mathbb{E}\left[\Pi\right]$ 
corresponds to averaging over the random instant $T$, distributed according to the pdf $f(t)$. 
The inner expectation is equal to
\begin{equation}
\mathbb{E}\left[\Pi\right] = \mathbb{E}\left[\Pi\left(T , [t_1,t_2]\right)\right] = \int_{t_1}^{t_2} f(t) dt ~.
\end{equation}
Using this last relation and the Poissonian statistics \eqref{rdempois} of the random number $R_d(m)$, 
the equality \eqref{gedexprgen} transforms into the Poissonian GPF:
\begin{equation}\label{gpfmeart}
G_d(z;[t_1,t_2]) =
\exp\left( \kappa\mu \cdot (z-1) \int_{t_1}^{t_2} f(t) dt \right) .
\end{equation}
In particular, if $t_1< t$ , $t_2>t$, then one may rewrite (\ref{gpfmeart}) as
\begin{equation}\label{gdtes}
G_d(z;[t_1,t_2]) = G_d(z;[t_1,t]) \cdot G_d(z;[t,t_2]) ,
\end{equation}
which is equivalent to the proposition.

\subsection{Proof of Proposition~\ref{propositiontwo}}

Before deriving equations \eqref{geaheqs}, it is useful to recall some properties of the total number of aftershocks
that are triggered by some shock, in the framework of the theory of (unmarked) branching processes. 
Each event triggers daughters (its first generation aftershocks), whose total number $R_d$ is described statistically 
by the following GPF,
\begin{equation}\label{goneprobs}
G_d(z) := \mathbb{E}\left[z^{R_d}\right] =
\sum_{r=0}^\infty q_d(r) \cdot z^r ~,
\end{equation}
where the $\{q_d(r)\}$ are the probabilities that the number $R_d$ of first-generation aftershocks is equal to a given integer $r$. In the framework of branching processes, all daughters trigger, independently of each other, their own daughters, whose numbers are iid random integers, possessing the same GPF $G_d(z)$, and so on.

Let $G_k(z)$ be the GPF of the number $R_k$ of the aftershocks of the first $k$ generations. Given the 
iid property of all numbers of any aftershock's daughters, we have
\begin{equation}\label{gekaplthruka}
G_{k+1}(z) = G_d\left[z G_k(z) \right] , \qquad k=1,2,\dots
\end{equation}
The product $z G_k(z)$ means that each daughter of the initial event triggers independently 
aftershocks belonging to the first $k$ generations, whose numbers are described by the same GPF $G_k(z)$.

A well-known result in the theory of branching processes states that, for $n\in(0,1]$ where $n$ is the
branching ratio defined as the average number of daughters of first-generation per mother,
\begin{equation}
n := \mathbb{E}\left[R_d\right] = \frac{d G_d(z)}{dz}\bigg|_{z=1}~ ,
\end{equation}
then the following limit exists
\begin{equation}\label{gkliminf}
\lim_{k\to\infty} G_k(z) = G(z)~ ,
\end{equation}
where $G(z)$ is the GPF of the total number of aftershocks of all generations that are triggered by the 
initial shock. Using the recurrent relation \eqref{gekaplthruka} and the limit \eqref{gkliminf}, $G(z)$ is
solution of the  transcendent equation:
\begin{equation}\label{gztranseq}
G(z) = G_d\left[z G(z) \right]~ .
\end{equation}

We can now derive the equation analogous to \eqref{gztranseq}, which determines the 
GPF $G(z;t)$ of the number $R(t)$ of future aftershocks of all generations. By future, we recall
that this refers to aftershocks that occur after the current time $t$, where the origin of time
is the time of occurrence of the main initial shock. Recall that the time intervals between a given
event and any of its directly triggered daughter are iid random variables with the same pdf $f(t)$.

We start with the derivation, similar to \eqref{gekaplthruka}, of the recurrent equation for the GPF $G_k(z;t)$ 
of the number $R_k(t)$ of triggered aftershocks of the
first $k$ generations. Let us discuss first the simplest case, where the shock has only one daughter (that is, $G_d(z)=z$), which is triggered at the random time $T$. Consider the conditional GPF $G_{k+1}(z;t|T)$ under 
the condition that $T$ is equal to the some given value.
In this case, the following relation holds
\begin{equation}\label{gkaplcas}
G_{k+1}(z;t|T) =
\begin{cases}
z G_k(z) , & T \geqslant t ,
\\
G_k(z;t-T) , & T < t .
\end{cases}
\end{equation}
Similarly to the right-hand side of equality \eqref{gekaplthruka}, the first line means that
the  GPF of the number of aftershocks of the first $k+1$ generations, including the shock's daughter and its aftershocks
of the first $k$ generations, is equal to $zG_k(z)$, This is because, if $T\geqslant t$, then the daughter and all its aftershocks are in the future (i.e. after $t$). In contrast, the second line of \eqref{gkaplcas} means that, if $T<t$, then the shock's daughter is not a future offspring, while we should only consider the future aftershocks of the daughter.

Let rewrite relation \eqref{gkaplcas} in the more convenient form for future analytical calculations:
\begin{equation}
G_{k+1}(z;t|T) = z G_d(z) \cdot \boldsymbol{1}(T-t) + G_k(z;t-T) \cdot \boldsymbol{1}(t-T) ~.
\end{equation}
Averaging both sides of this equality with respect to the statistics of the random time $T$ with pdf $f(t)$, we obtain
\begin{equation}\label{gekaztlinver}
G_{k+1}(z;t) = z G_k(z) \Phi(t) + \int_0^t f(\tau) G_k(z;t-\tau) d\tau~ .
\end{equation}

Let us now get the sought recurrent equation in the general case where the GPF  $G_d(z)$ of the number of daughters is arbitrary. Since the time durations $\{T_k\}$ between any shock and its first-generation daughters are iid variables,
in order to obtain the recurrent equation, one needs to replace in \eqref{gekaplthruka} the GPF $G_{k+1}(z)$ by $G_{k+1}(z;t)$, and $z G_k(z)$ by the right-hand side of the equality \eqref{gekaztlinver}, that is to say
\begin{equation}\label{gekaplthrukazt}
G_{k+1}(zæå) = G_d\left[z G_k(z) \cdot \Phi(t) + f(t) \otimes G_k(z;t) \right] , \qquad k=1,2,\dots
\end{equation}
For $n\in(0,1]$, then a limit similar to \eqref{gkliminf} holds
\begin{equation}
\lim_{k\to\infty} G_k(z;t) = G(z;t)~ .
\end{equation}
In this limit, we obtain from \eqref{gekaplthrukazt} the sought equation for the GPF $G(z;t)$ of the number 
of future aftershocks of all generations:
\begin{equation}\label{gkliminfte}
G(z;t) = G_d\left[z G(z) \Phi(t) + f(t) \otimes G(z;t) \right] ~.
\end{equation}
It is easy to check that this equation (\ref{gkliminfte}) is equivalent to \eqref{geaheqs}.

\subsection{Proof of proposition~\ref{propositionthree}}

After substitution relations \eqref{thetoutinexpr} into the right-hand side of equality \eqref{Thetasplit}, 
we obtain expression \eqref{thetztevent} for the GPF $\Omega(z;t,n,m)$ of the number of future aftershocks
of all generations. In turn, taking into account the second equation in \eqref{geaheqs} and equality \eqref{qydef}, 
we obtain equation \eqref{hcloseq}.

\subsection{Proof of proposition~\ref{propositionfour}}

By definition, the GPF $\Omega(z;t,n)$ of the number of future aftershocks of all generations is equal to
\begin{equation}\label{thetztnqu}
\Omega(z;t,n,m) := \mathbb{E}\left[z^{R(t)}\right] = \sum_{r=0}^\infty q(r;t,n,m) \cdot z^r ,
\end{equation}
where $R(t)$ is the random number of the future aftershocks of all generations, while $\{q(r;t,n,m)\}$ are the probabilities that the random number $R(t)$ is equal to the given integer $r$.

Let $R_\nu(t)$ be the random number of future offsprings whose magnitudes exceed the threshold $\nu$,
\begin{equation}\label{rtmoudef}
R(t;\nu) = \sum_{j=1}^{R(t)} \boldsymbol{1}(m_j-\nu) ,
\end{equation}
where $\{m_j\}$ are the magnitudes of the future offsprings.

Using the law of total probability, we can represent the GPF of the random number $R_\nu(t)$  in analogy with equality \eqref{thetztnqu} under the form:
\begin{equation}\label{thetztmmdef}
\Omega(z;t,n,m,\nu) := \mathbb{E}\left[z^{R(t;\nu)}\right] = \sum_{r=0}^\infty q(r;t,n,m) \cdot \mathbb{E}\left[z^{R_\nu(t)}|r\right] ~,
\end{equation}
where $\mathbb{E}\left[\cdots | r\right]$ is the conditioned expectation, under the condition that the number of future offsprings 
is equal to the given integer $r$: $R(t)=r$.

Taking into account that, in the framework of the ETAS model, all offsprings have iid random magnitudes that are statistically independent of the number $R(t)$ of future aftershocks, we obtain
\begin{equation}
\mathbb{E}\left[z^{R_\nu(t)}|r\right] = \Lambda^r(z,\nu) , \qquad \Lambda(z,\nu) := \mathbb{E}\left[ z^{\boldsymbol{1}(m'-\nu)}\right] ~,
\end{equation}
where $m'$ is the random magnitude of some offspring distributed according to the GR law \eqref{grldef}. Using the identity
\begin{equation}
z^{\boldsymbol{1}(m-\nu)} \equiv  1 + (z-1) \cdot \boldsymbol{1}(m-\nu) ~,
\end{equation}
similar to \eqref{zetpopi}, we obtain
\begin{equation}
\Lambda(z,\nu) = 1 + (z-1) \cdot p(\nu) \quad \Rightarrow \quad \mathbb{E}\left[z^{R_\nu(t)}|r\right] = \left[1 + (z-1) \cdot p(\nu) \right]^r ,
\end{equation}
After substitution the last relation into \eqref{thetztmmdef}, and after performed the summation of the series, we obtain the sought relation \eqref{thetaemou}.

\end{document}